\documentclass[]{iopart}
\usepackage{epsfig}
\usepackage{fancyhdr}
\usepackage{graphicx}
\usepackage{latexsym}
\usepackage{amssymb}

\newcommand{\intbeta}{\beta^\mathrm{int}_e}
\newcommand{\intte}{T^\mathrm{int}_e}
\newcommand{\inttp}{T^\mathrm{int}_p}
\newcommand{\gyro}{\Omega_e^\ast}
\newcommand{\gyrotime}{\gyro t}
\newcommand{\inertial}{c/\omega_e}
\newcommand{\ep}{$\mathrm{e}^{-}\!\!\!-\!\mathrm{p}^{+}$}
\newcommand{\ee}{$\mathrm{e}^{-}\!\!\!-\!\mathrm{e}^{+}$}

\begin{document}

\title[On particle acceleration and trapping by PFDO's]
{On particle acceleration and trapping by Poynting flux dominated flows}
\author{Gunnar  Paesold\dag, Eric G. Blackman\dag\ddag\ and Peter Messmer\P}
\address{\dag\ Department of Physics \& Astronomy, University of Rochester,
Rochester, NY 14627, USA}

\address{\ddag\ Laboratory for Laser Energetics, University
of Rochester, Rochester, NY 14627, USA}
\address{\P\ Tech-X Corporation, 5621 Arapahoe Avenue, Boulder, CO 80303, USA}
\ead{paesold@pas.rochester.edu, blackman@pas.rochester.edu}

\label{firstpage}

\begin{abstract}
Using particle-in-cell (PIC) simulations, 
we study the evolution of a strongly magnetized 
plasma slab propagating into a finite density ambient medium. 
Like previous work, we find that the slab breaks 
into discrete magnetic pulses. The subsequent evolution is consistent
with diamagnetic relativistic pulse
acceleration of~\cite{liangetal2003}.
Unlike previous work, we use the actual electron to proton mass ratio
and focus on understanding  trapping 
vs. transmission of the ambient plasma by the pulses and on the 
particle acceleration spectra. We find that the accelerated electron
distribution internal to the slab develops a double-power law.
We predict that emission from reflected/trapped external electrons
will peak after that of the internal electrons.
We also find that the thin discrete pulses trap ambient electrons
but allow protons to pass through, resulting in less drag on the pulse
than in the case of trapping of both species.  
Poynting flux dominated scenarios have been proposed
as the driver of relativistic outflows and particle
acceleration in the most powerful astrophysical jets.
\bigskip

(IN PRESS, PLASMA PHYSICS AND CONTROLLED FUSION)
\end{abstract}

\maketitle
\section{Introduction}
\label{intro}
Relativistic plasma outflows occur in 
the most powerful sources in the universe.
The inferred Lorentz factors vary from 
$\sim 10^1$ in active galactic nuclei (AGN) to $> 10^2$  
in gamma-ray bursts (GRB). 

In AGN, it is widely believed that magnetic fields
play a key role in launching and driving the observed jets,
but how far away from the source (e.g. a few engine radii vs. 
parsec scales) the flows remain Poynting flux dominated is unresolved. 
A related debate for GRB is
whether  the relativistic outflows are powered by hot matter dominated
outflows (MDO) or cold Poynting flux dominated outflows (PFDOs).
In the latter, the energy
is carried electromagnetically to great distances where it is finally
converted to particle energy.
For GRB, MDO's are constrained by the fact that too much 
baryon loading prevents the flow  from reaching the large Lorentz
factors needed to overcome the compactness
problem~\cite{piran1999}. The total energy density must greatly
exceed the rest mass energy density of the outflow. 
In contrast PDFO's can in principle
transport large amounts of energy through a vacuum without much matter. 
Poynting flux driving is a part of a number of models: 
tori in neutron star mergers~\cite{narayanetal1992, thompson1994,
meszarosrees1997,katz1997},  highly magnetized millisecond
pulsars~\cite{usov1992,byf96,spruit1999}, and 
collapsars~(e.g.~\cite{mwh01,wmw02}). See also the recent comprehensive
study of~\cite{2003astro.ph.12347L}.


For PFDOs, the particle acceleration occurs 
from dissipation of the magnetic energy.
Understanding the physics thereof requires studying relativistic
collisionless plasmas. 
Several  mechanisms of PFDO dissipation 
have been proposed in the literature. 
Direct dissipation of magnetic energy by fast reconnection 
has been considered~\cite{lb01, spruitetal2001, drenkhahn2002,
lyutikovuzdensky2003}. Alternatively, Ref. ~\cite{smolskyusov1996} modeled
the interaction of a strongly magnetized wind
with an external medium. In the 
wind rest frame this scenario is identical to the collision of a wide
relativistic cold beam of particles with a strong 
magnetic barrier. The particle acceleration is driven by
electromagnetic fields induced from charge separation 
between protons and electrons dragged from the ambient medium
by the barrier.
Electrons can acquire a substantial fraction of the proton kinetic
energy to produce synchrotron emission.  

More recently, a new but related acceleration
mechanism has been proposed by~\cite{liangetal2003} 
involving a hot magnetized collisionless plasma, confined to a 
finite slab (rather than the infinite slab of~\cite{smolskyusov1996})
and suddenly released into a vacuum.  The surface gradient of
the magnetic field generates a strong diamagnetic current which
shields the interior and confines the field to the expanding
plasma, but the initial magnetic slab breaks into multiple 
pulses. An electric field of order $|E|\sim c|B|$ is associated with
the magnetic field  of each pulse. 
Particles are trapped in the pulse surface layer and  accelerated  to higher 
and higher energies via the ponderomotive force of the 
magnetic pressure gradient. This mechanism has been termed
the diamagnetic relativistic pulse accelerator (DRPA) by~\cite{liangetal2003}.

The wind set-up~\cite{smolskyusov1996}
can  be thought of as a special
case of a magneto-acoustic pulse in the fnite pulse model~\cite{liangetal2003}
for which the pulse has an infinite width in the propagation
direction.  We will demonstrate that the pulse width
is particularly important for the interaction with the ambient medium.
Depending on whether the width is larger or smaller than the particle
gyro-radii there are three general possibilities: {(\em i)} the
external matter crosses the pulse 
and is lost to the interior of the expanding plasma, {(\em ii)} the
external particles get trapped and accelerated inside the pulse or
{(\em iii)} the external particles are reflected into the oncoming material.
(In the relativistic regime, we will see that  
{(\em ii)} and {(\em iii)} are largely indistinguishable.)
For realistic pulses, 
the rise of the magnetic field strength at the leading edge of the leading 
pulse will be smoother  than the abrupt jump assumed  
in~\cite{smolskyusov1996} which makes a significant difference.
Complementarily, although an ambient medium
was included in~\cite{liangnishimura2004} its effect 
not yet been thoroughly studied in detail.
That is a goal of this work.

In addition to investigating the  interaction between
the pulse and the ambient medium, we will also
study the spectrum of the both the internal and external particles.
While many of the particles initially inside the pulse remain trapped,
they get accelerated to high energies. This fast population of electrons
will emit synchrotron radiation.
The interaction of the pulse
with the ambient medium can deliver a second delayed contribution
to the  emission:
In cases where the external material gets trapped and accelerated in
the pulse,  (case {\em (ii)}  above),  it is expected that these
captured electrons also radiate  synchrotron
emission. This emission can be delayed with respect to the primary
prompt emission 
if the density of the ambient medium is low compared to the density of
the internal matter: a significant time is then needed for electrons
to pile up to densities that produce 
observable radiation. 
This secondary emission can either blend in with the prompt emission
and determine the late evolution thereofore can be seperated in time
and form a completely distinct part of the systems radiation emission. 

We tackle the problem of magnetic pulse-ambient medium interaction
with fully relativistic PIC simulations. The magnetic pulse width,
shape and strength are not assumed, but are generated
self-consistently from the driving event, namely  
the sudden expansion of the strongly magnetized  plasma. 
Unlike previous simulations, we use the real electron
proton mass ratio of $m_e/m_p=1/1836$ (see e.g.~\cite{liangetal2003}).

In Sec.~\ref{sim} we describe the 
code used and the simulation set-up. Sec.~\ref{interaction}
addresses the basic mechanism of particle reflection at a magnetic
pulse and a simple model is presented and compared to simulation
results. The long term pulse evolution is discussed 
in Sec.~\ref{longterm}. We discuss the accelerated particle spectrum
in Sec.~\ref{radiation}. 
In Sec.~\ref{filter} we show that the pulses
can act as electron selective filters. In Sec.~\ref{application} a
possible application to GRBs is discussed. We summarize in
Sec.~\ref{summary}. 
\section{Simulation Setup and Code}
\label{sim}
\begin{figure}
  \begin{center}
  \resizebox{0.7\hsize}{!}{\epsfig{file=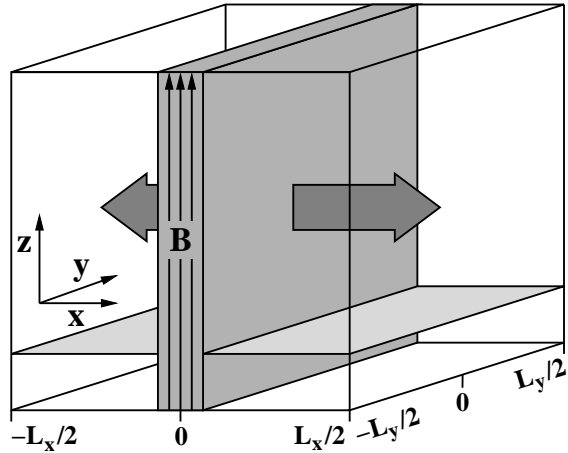}}

  \end{center}
  \caption{Sketch of the simulation box setup. The initial internal
    plasma is contained in the central plasma slab which is permeated
    by a magnetic field. The volume outside the slab is filled with
    the ISM (shown as transparent for clarity). The
    light shaded plane is the simulated 2D portion of the
    volume and the arrows indicate the direction of expansion.}
  \label{fig:1}
\end{figure}

The basic physical setup is as follows: an initial magnetic field in
the z-direction, i.e. $\vec{B}_0=(0,0,B_z)$, is placed in a plasma
slab of finite thickness at $x=0$.  
Due to the magnetic pressure gradient, the
slab expands in both the positive and negative x-directions (see
 Fig.~\ref{fig:1}). 
As in~\cite{liangetal2003} we focus on the part of the slab expanding 
into the  the $+x$-direction. 
Throughout this work, times are normalized to
the initial non-relativistic gyro frequency of the internal electrons
$\Omega_e^\ast=eB_0/m_e$. 

The code VORPAL~\cite{nietercary2004} used in our simulations is a
fully relativistic 3D open source plasma simulation code based on the
PIC algorithm 
presented in ~\cite{birdsalllangdon1991}. VORPAL can be used in
an arbitrary number of phase space dimensions.  
The simulations herein are referred to as $2\frac{1}{2}$
dimensional, referring to two dimensions in space (see
Fig.~\ref{fig:1}) and three in velocity 
space. The code is freely available for academic use under a public
license agreement. Our simulations are performed on a linux cluster
with 8 CPUs dedicated to the computations. We were therefore able to
extend single simulation runs over months, resulting in simulation times
much longer than in previous analyses.  

The PIC algorithm
introduces a uniform spatial grid to calculate field
quantities. The particles' space and velocity coordinates 
are continuous, and advance through extrapolation of the field quantities
and the resulting forces. The grid spacing throughout our 
simulations is given by the inertial length of the initial electrons.
$\Delta_x=\Delta_y=\Delta=c/\omega_e\approx0.0168\cdot(n^{int}_{e,0}/10^{17}\;
\mathrm{m^{-3}})^{-1/2}$m, where 
$\omega_e=\sqrt{e^2n^{int}_{e,0}/\epsilon_0m_e}$ is the electron plasma frequency
defined in the laboratory frame, and $n^{int}_{e,0}$ is the initial internal
electron density.
The $x$ and $y$ domains are periodic and run from $-L_x/2$ to $+L_x/2$ and 
$-L_y/2$ to $+L_y/2$ respectively.  The magnetic slab initially centered
at $x=0$,   expands in the $x$-direction. Since a magnetic pulse
propagates at approximately $c$, the simulation time-spans are given
by the time for light to cross half of the box
in the x-direction, i.e. $t\approx(L_x/2)/c$, ensuring that the pulse does
not leave the box. The time resolution of the
simulation is chosen such that  
$\Delta t\approx 0.1\cdot\min{[1/\Omega_e,1/\omega_e]}$. All
quantities are presented in dimensionless units. 
The choice of one parameter, e.g. the
initial magnetic field, then fixes all other quantities for a given
setup. Throughout the following,
physical quantities are calculated 
for an initial magnetic field of $B_0=4.473\;\mathrm{T}$.  

\begin{figure}
  \begin{center}
     \resizebox{0.7\hsize}{!}{\epsfig{file=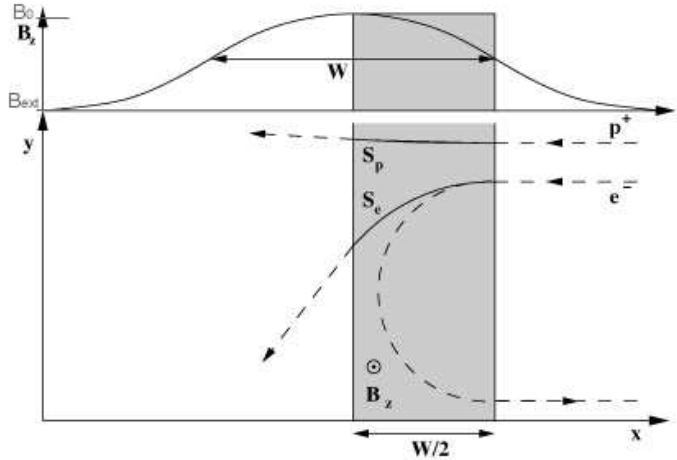}}
  \end{center}
  \caption{Sketch of external electron and proton 
    motion in the pulse rest frame. The pulse we study is the leading 
pulse that breaks off from the initial slab.
 Depending on the pulse width $W$ and its
    velocity, an incoming electron can either be reflected or
    transmitted. For the parameter ranges considered, protons are always
    transmitted. $S_p$ and $S_e$ are the path lengths inside the first
    half of the pulse for protons and electrons respectively. Only the leading
 half of the pulse (shaded) matters for assessing transmission or reflection.
The magnetic field range is shown from 
 $B=B_{ext}$ at the base of the Gaussian-like field profile,
to $B=B_{0}$ at the apex.}

  \label{fig:2}
\end{figure}
Plasma particles are treated as macro particles, each corresponding to
many real particles. In the simulations herein, each
particle species is  initially represented by 50 macro particles per
cell. Typically millions of particles are simulated. The particle
species are indicated by subscripts with  
the subscript $int$ and $ext$ indicating particles initially 
inside or outside of the magnetic slab. 
These internal and external particles can be further subdivided
into electron and ion or electron and positron populations. 
In the context of a GRB, the external plasma 
would represent the interstellar medium (ISM).  
For the external plasma we use $|B_{ext}|=10^{-10}\;\mathrm{T},
n_e^{ext}=n_p^{ext}=10^6\;\mathrm{{m}^{-3}}$. 

\begin{figure}
  \begin{center}
     \resizebox{0.6\hsize}{!}{\epsfig{file=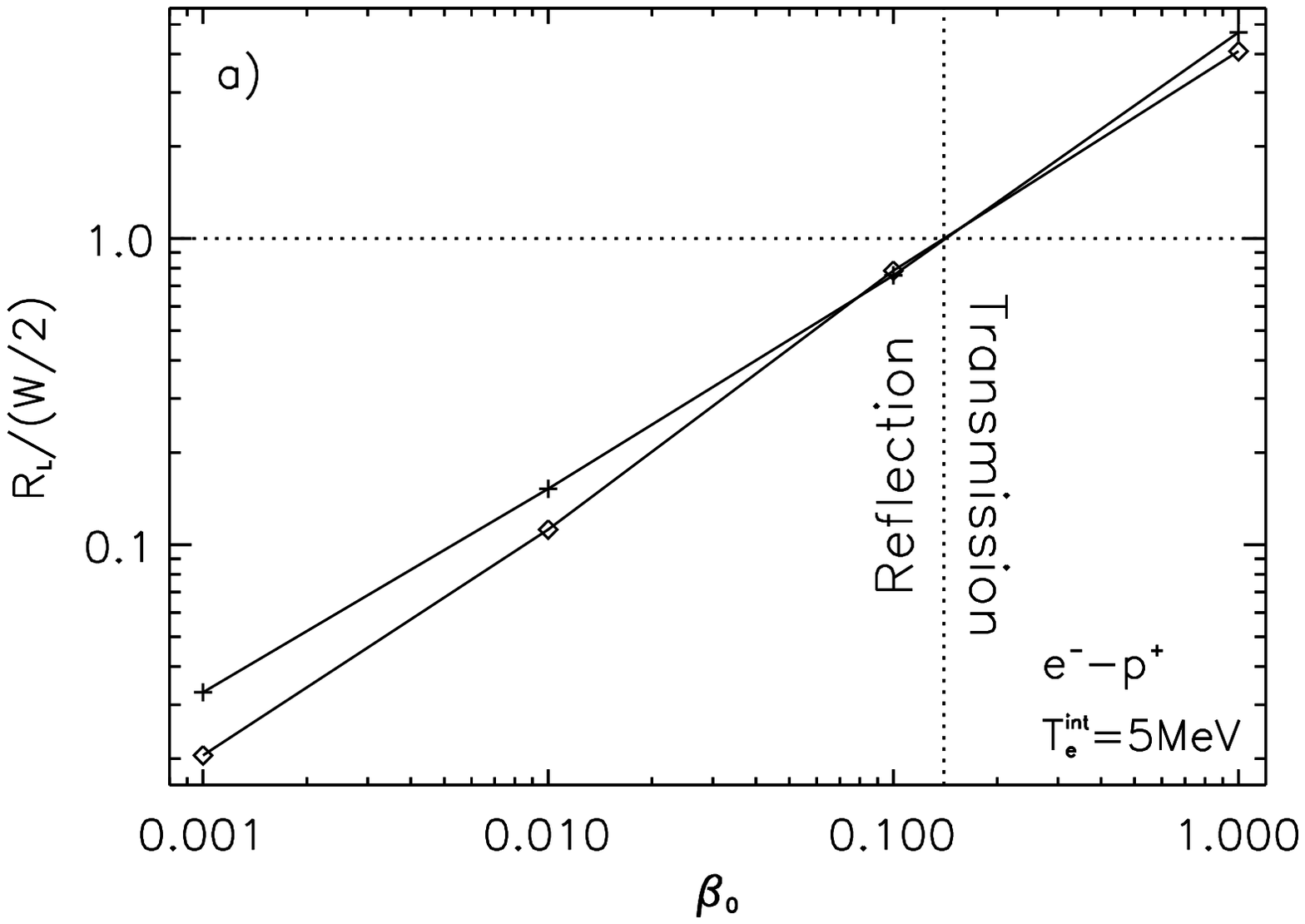}}
     \resizebox{0.6\hsize}{!}{\epsfig{file=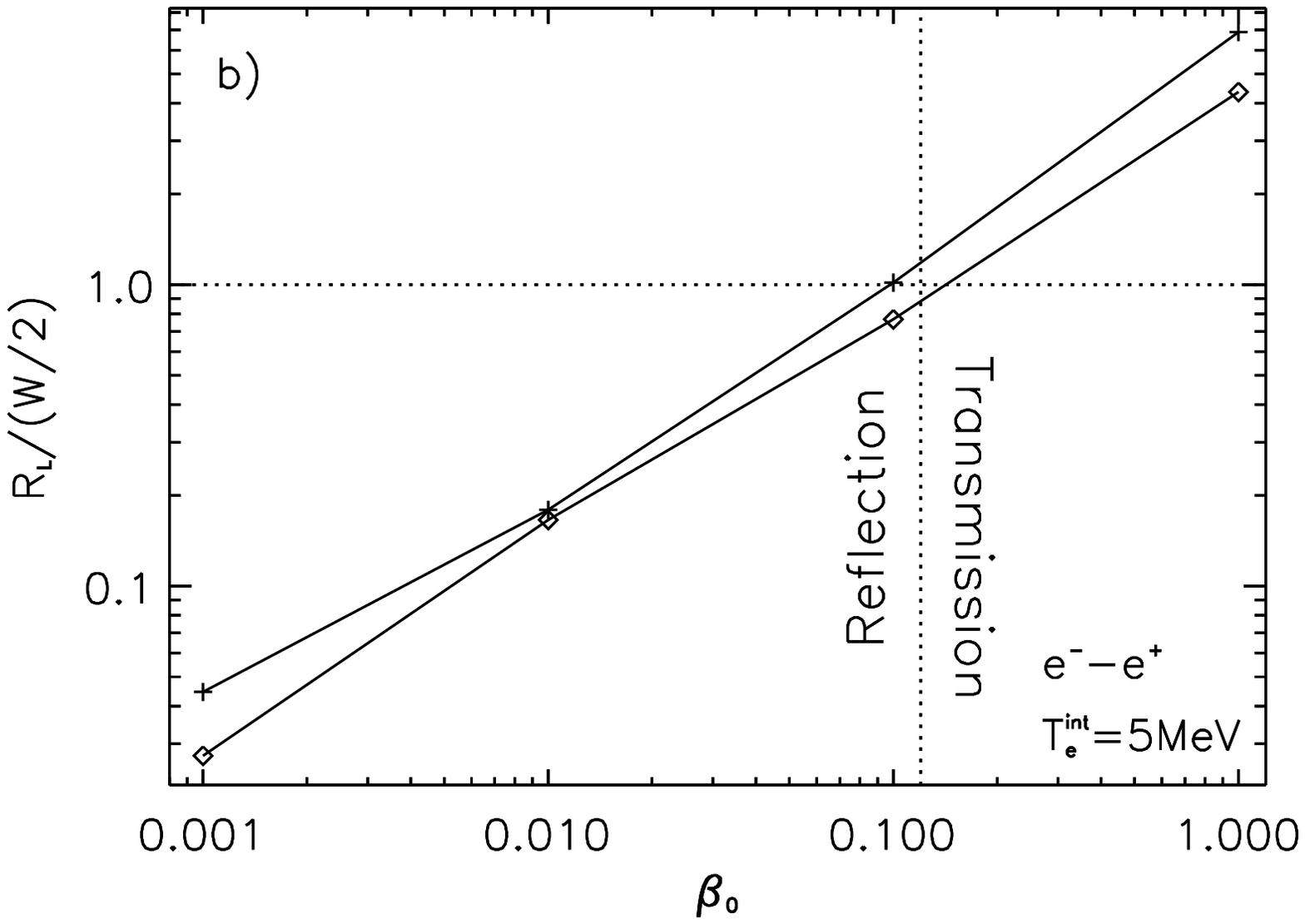}}
     \resizebox{0.6\hsize}{!}{\epsfig{file=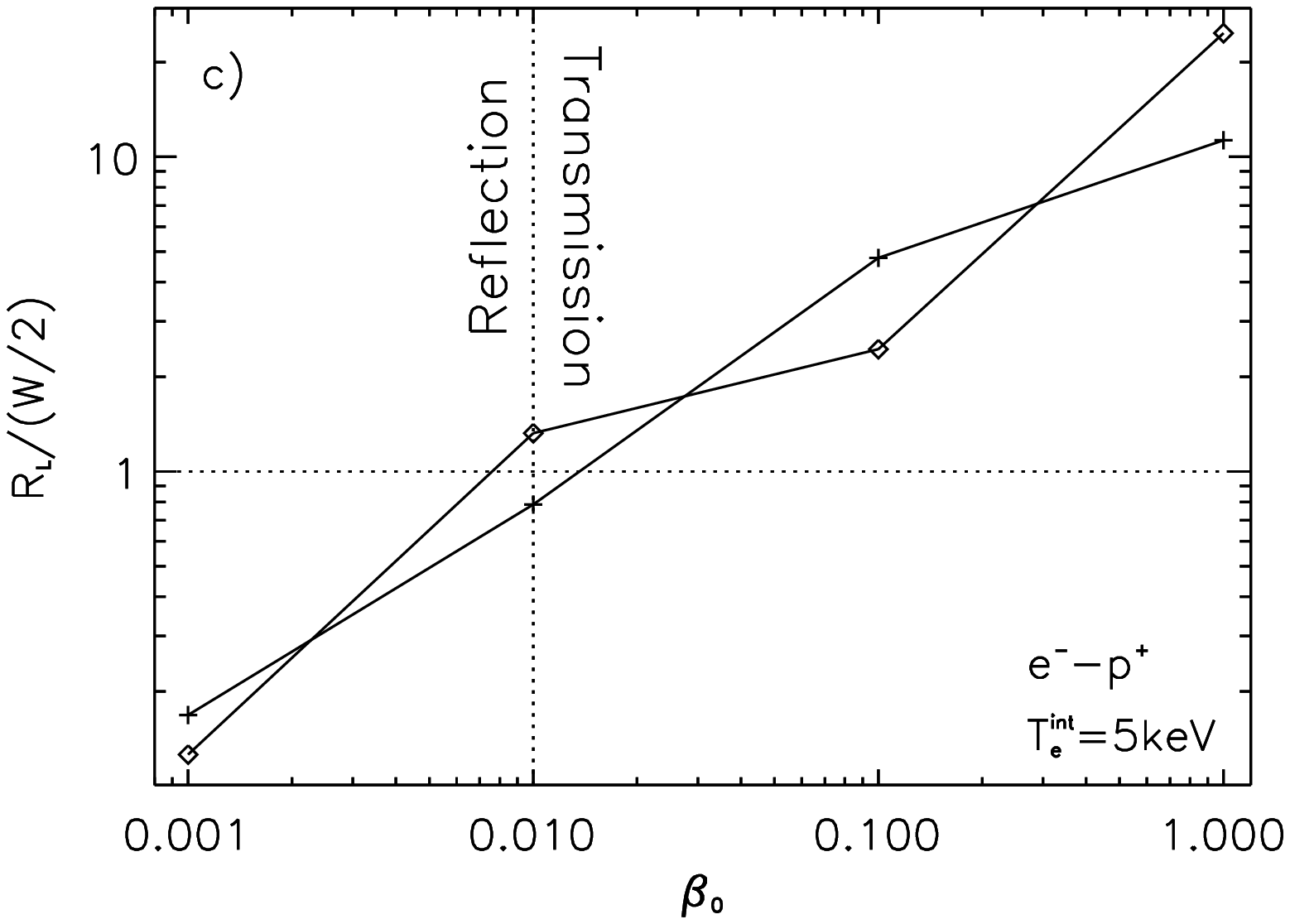}}
  \end{center}
  \caption[]{Ratio of Larmor radius $R_L$ of the external electrons
                            inside the pulse to 
                            $W/2$, the  HWHM value of the pulse measured
                            in the pulse rest frame. The plots a) and
                            b) show the simulation results for an
                            initial 
                            temperature of $5\;\mathrm{MeV}$ for an
                            \ep and an \ee internal plasma
                            while plot c) shows results for an \ee internal plasma with
                            initial temperature of
                            $5\;\mathrm{keV}$ (open squares are 
                            lower density case; crosses are higher
                            density case).} 
  \label{fig:3}
\end{figure}
Variation of the internal plasma parameters is described below, 
 but particles are always initially loaded with
 relativistic Maxwellian distribution of
temperature $T$ throughout the 
simulation volume. Unlike previous authors who invoked 
 a reduced mass ratio $m_p/m_e=100$ ~\cite{nishimuraetal2003}, 
we use the actual ratio of $m_p/m_e=1836$.  
\section{Particle transmission, reflection, and trapping}
\label{interaction}
\begin{figure}
  \begin{center}
     \resizebox{\hsize}{!}{\epsfig{file=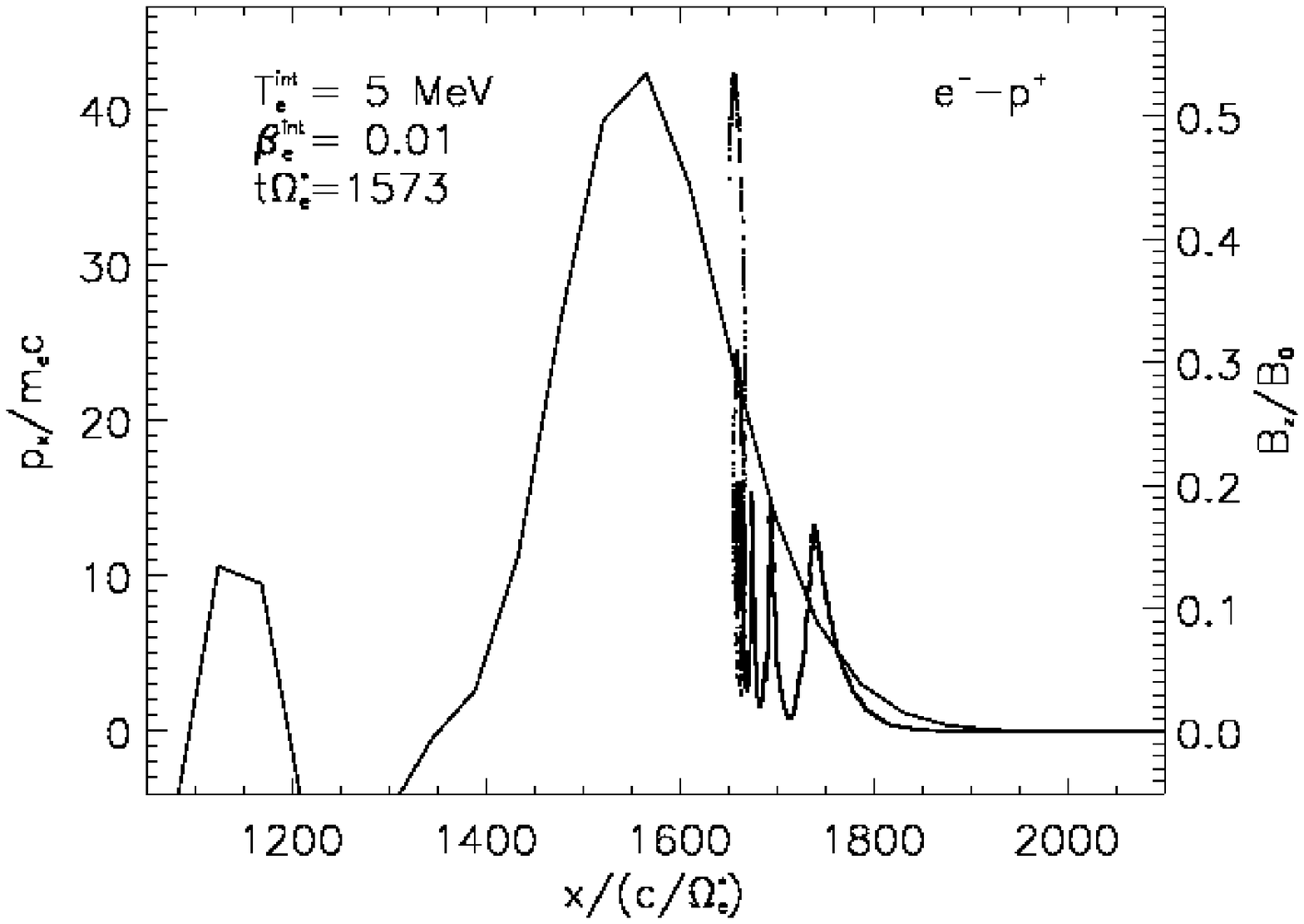}
                           \epsfig{file=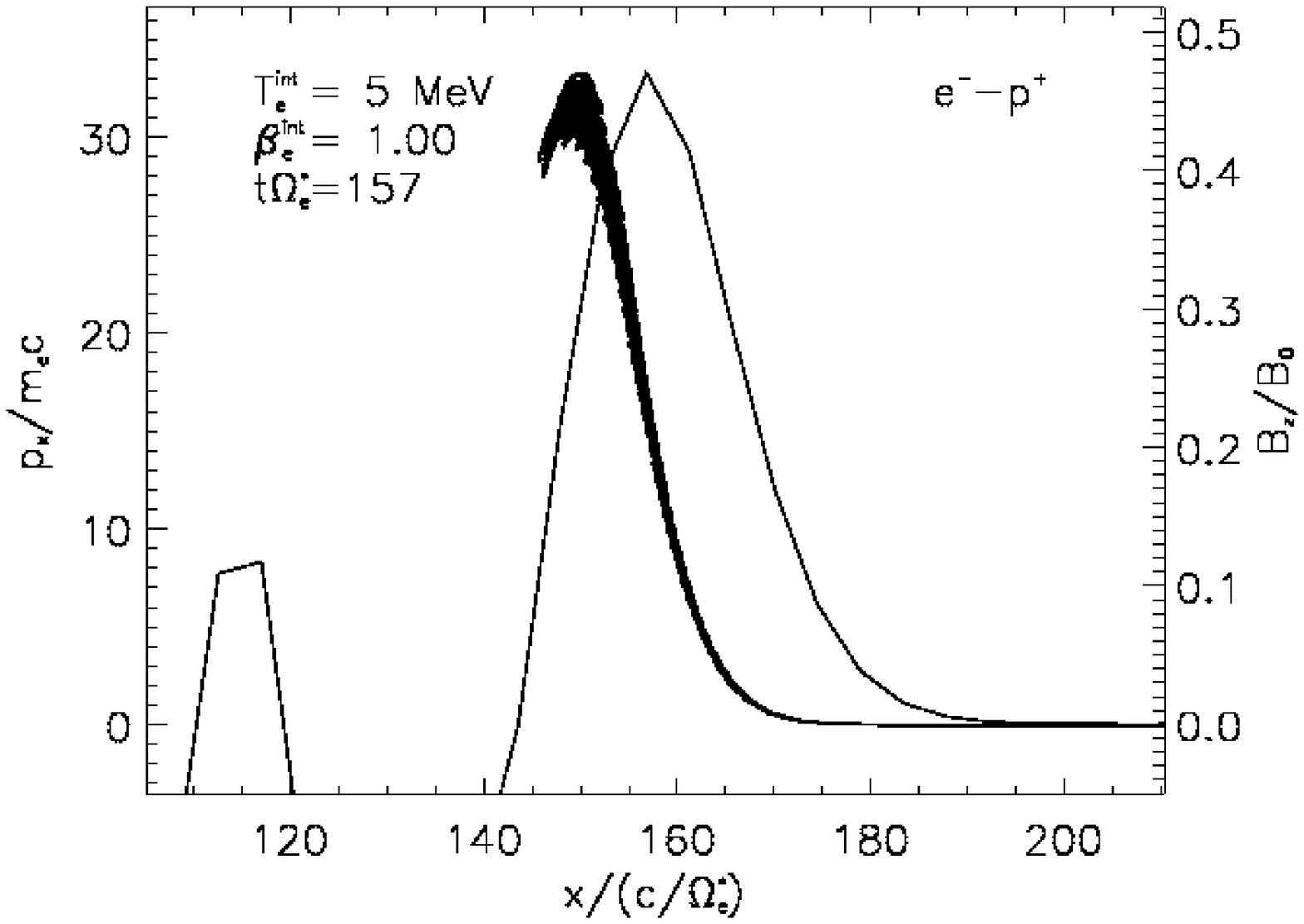}}
     \caption[]{Phase space plot of the external electrons with
                over plotted magnetic field 
                profile in the laboratory
                frame. Electron-proton plasmas at a
                temperature of 
                $T_e=T_p=5\mathrm{MeV}$ are displayed. On
                the left hand side the initial internal plasma
                beta is $\beta_e^\mathrm{int}=0.01$, a value at
                which electron reflection is expected
                according to Fig.~\ref{fig:3}. On the
                right hand side an initial plasma beta of
                $\beta_e^\mathrm{int}=1.0$ is depicted. The
                times in the evolution of the system
                have been chosen to be equal in real time.} 
  \label{fig:4}
  \end{center}
\end{figure}
Including a finite density ambient medium 
in the DRPA process allows the 
interaction between the slab and ambient medium to be studied.
Early simulations with 
$\beta_\mathrm{int}\geqslant0.1$ (where $\beta = 2
\mu_0
n_ek_BT_e/B_0^2$), suggested that external material is
neither reflected nor piled up ahead of the pulse, whereas simulations with
initially low $\beta$ show a pile-up of external electrons. Due to the
high degree of relativity, reflection and trapping of external
particles are not
distinguishable in the lab frame. Particles cannot
propagate ahead of the pulse significantly, and
stay close to its front like trapped particles. Reflection 
and trapping will thus be used synonymously below.
\subsection{Transmission vs. reflection: the basic picture}
\label{basic}
Our simulations are performed in the laboratory frame where the mean
initial velocity of all particle species vanishes. In order to analyze
the conditions for transmission vs. reflection of external material it
is convenient to transform the simulation results into the pulse 
rest frame. Below, 
un-primed quantities refer to values measured in the rest frame and
primed quantities are in the laboratory frame. 
%
%

From Ohm's law, the perpendicular electric field in the pulse vanishes in
its rest frame.
External particles penetrate the pulse as a cold beam with speed
$v_\mathrm{ext}=-v_\mathrm{pulse}$. The relativistic
gyro-frequency for an incoming particle of species $\alpha$ in the pulse
rest frame is therefore given by
\begin{eqnarray}
    \Omega_\alpha^\mathrm{ext}=\frac{q_\alpha B}
    {m_\alpha\Gamma_\mathrm{pulse}c}\quad,
\end{eqnarray}
where $\Gamma_\mathrm{pulse}=1/\sqrt{1-v^2_\mathrm{pulse}/c^2}$ is the
Lorentz factor of the pulse in the laboratory frame. The pulse magnetic
field in the rest frame is $B=B^\prime/\Gamma_\mathrm{pulse}$, where
$B^\prime$ is the magnetic field in the laboratory frame.

Figure~\ref{fig:2} shows the situation in the  pulse frame. 
In assessing whether 
or not a particle is reflected, only its trajectory in 
the leading half of the pulse matters.
Analyzing whether a particle will reflect is 
simplified by approximating the leading half of the pulse profile by a
magnetic field filling a rectangular area of width $W/2$ (gray shaded
area in Fig.~\ref{fig:2}). As seen from the sketch, 
reflection requires the half-pulse width to exceed the Larmor radius.

Ref. \cite{nishimuraliang2004} showed that 
the width $W$ of the leading pulse 
that breaks off from the slab 
 scales as the width of the initial slab. 
Fig.~\ref{fig:2} shows  that the larger the pulse width is 
compared to the external particle's gyro-radius,  
the more likely that particle will reflect. Ref. (\cite{smolskyusov1996}
considered an essentially infinite pulse, and accordingly,  
particles are all reflected).

The pulse shape is not maintained over long
times~\cite{liangnishimura2004} and instead bifurcates   
into  a complex structure of multiple peaks. The sub-pulses 
have widths of order $\sim 10 c/\omega_{e}$ 
and are otherwise independent of the initial plasma properties. 
Even if 
the leading pulse is too large to allow particle transmission,
the ultimate reflection or transmission is determined by 
the behavior of smaller sub-pulses in a later
phase of the 
pulse evolution. Because computational limits
are prohibitive, early development of small pulses
must be created by choosing a small enough initial plasma slab.
\subsection{Numerical study of parameters influencing  reflection
  vs. transmission} 
The results of our simulations 
confirming the basic condition for  reflection are 
shown in Fig.~\ref{fig:3}.
We used two different temperatures
$T_e^\mathrm{int}=T_p^\mathrm{int}=5\;\mathrm{keV},5\;\mathrm{MeV}$
and four values of the internal electron-to-magnetic pressure ratio
$\beta_e^\mathrm{int}
=0.001,0.01,0.1,1.0$.  The internal density was varied
by a factor of four for for each
choice of $\beta_e^\mathrm{int}$. The width of the initial
slab $W_s$ is held constant between runs so such that 
$W_s =6\;c/\omega_e^{low}=12\;c/\omega_e^{high}$ in grid spacing units,
where 'high' and 'low' refer to the corresponding density cases. 
We used an initial external plasma composition of \ep and 
internal plasma compositions of either \ep or \ee.


The simulation box dimensions are $L_x=500\;c/\omega_e$
and $L_y=80\;c/\omega_e$. The initial plasma beta is
$\beta_e^\mathrm{int}=0.01$ and the simulations cover a timespan of
$\gyrotime\approx11025$. 

By measuring the lab frame pulse velocity, field strength,
and width, we obtain the corresponding values in the pulse
rest frame. Comparison of the electron Larmor radius with the pulse
width yields the data of Fig.~\ref{fig:3}. The results of
the \ep and the \ee simulations at temperatures
of $5\;\mathrm{MeV}$ and $5\;\mathrm{keV}$ are shown. Values of
$R_L/(W/2)<1$ indicate that 
electrons should be reflected while $R_L/(W/2)>1$ show the
transmission of external electrons to the internal material
downstream of the pulse. 
As can be seen from the \ep simulations, 
the transition value roughly scales with the plasma temperature. 

The high temperature \ee results are as clean as the \ep 
case. The critical $R_L$ marking the transition 
from reflection to transmission is the same as in the \ep case.
In contrast,  the transition between reflection and transmission 
in the low temperature \ee simulations is hard to pinpoint because
the pulse structure is more quickly influenced by 
the pick-up of  external electrons. Counter streaming currents 
perpendicular to the pulse velocity drive instabilities, causing the
pulse to fragment early. For an \ep plasma, trapping of 
electrons has much less effect on the overall pulse
structure on the same time-scales.

The initial plasma density has little influence on the results in
Fig.~\ref{fig:3}a-c.  The transition value of
the initial $\beta_0$ therefore scales primarily 
with the plasma temperature.  The lower the temperature the
lower the limiting $\beta_0$.

The linear dependence of $R_L/(W/2)$ on $\beta_0$ as shown in
Fig.~\ref{fig:3}a-c is not strictly valid
on all scales. Extending the analysis to very small magnetic field
strengths leads to a break down of the linearity. In the case of large magnetic
fields, however, the linear dependence is observed in all simulations.
More work is needed to systematically understand the dependence for
much larger and much smaller field strengths.

The observed linear dependence of $R_L/(W/2)$ on $\beta_0$ allows us
to derive a scaling for the pulse  Lorentz factor
$\Gamma_\mathrm{Pulse}$ in the present parameter region. Using
$W\propto c/\omega_e$ and the definition of $R_L$ with
$R_L/(W/2)\propto \beta_0$ we find the following scaling of the pulse
Lorentz factor 
\begin{eqnarray}
\Gamma_\mathrm{Pulse}\propto \frac{n_e^{1/2}T_e}{B_0}\quad.
\end{eqnarray}
This equation is valid under the assumption that the density of the
ambient medium is small enough not to influence the pulse propagation.

\subsection{Phase space analysis of trapping/reflection}
Figure~\ref{fig:4}  illustrates the differences in pulse--particle
interaction for the cases of transmission vs. trapping: Fig.~\ref{fig:4}a shows the external electron phase
plot $x-p_x$ superimposed on the magnetic
field profile of the pulse in the 
laboratory frame for an initial $\beta_e^\mathrm{int}=0.01$,
a value for which reflection/trapping is expected. 
As shown, the external electrons  pile up at the 
pulse. In contrast, Fig.~\ref{fig:4}b shows
the analogous plot for $\beta_e^\mathrm{int}=1.0$. 
Here transmission is expected and seen.

\begin{figure}
  \begin{center}
     \resizebox{0.7\hsize}{!}{\epsfig{file=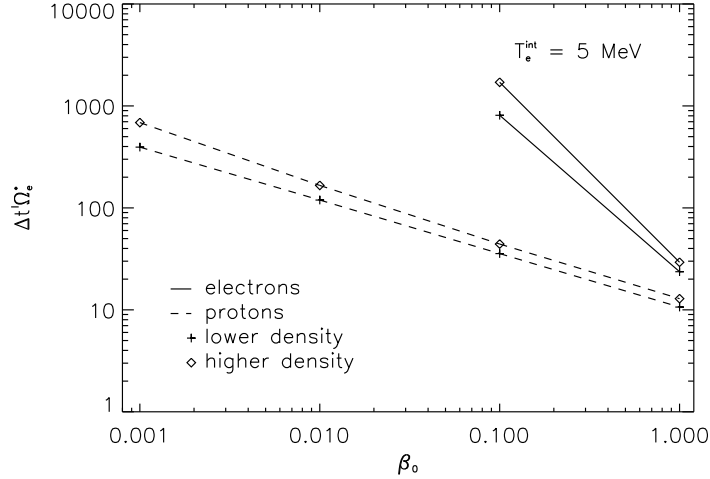}}
  \end{center}
  \caption[]{Time needed for electrons and protons to cross the pulse
    in the laboratory frame. The electron solutions do not exist for
    $\beta_\mathrm{e}< 0.1$ since $R_L^e < W/2$ and the particles cannot
    cross the pulse. Since $R_L^p\gg W/2$, the proton solutions in this
    regime correspond to the pulse crossing time for a free flying
    particle.} 
  \label{fig:5}
\end{figure}
\subsection{Electron and  proton pulse crossing times}
Electrons take longer to cross the pulse
than the protons.  To show this,   
we first define  the pulse crossing time as that needed
to  travel half the pulse width $W/2$. In the rest frame, 
the particle propagates 
in  the $-x$ direction with Lorentz factor
$\Gamma_\mathrm{pulse}=1/\sqrt{1-(v_\mathrm{pulse}/c)^2}$. 
The space-time event of the pulse encounter occurs at $(t_1,x_1)$ and 
the event of exit is at 
$(t_2,x_2)=(t_1+\Delta t,x_1-W/2)$ (Note: $x_2$ is
given by $x_2 = x_1 - W/2$), where $\Delta 
t$ is the time spent inside the pulse. 
The time interval in the laboratory frame is then given by 
\begin{eqnarray}
\Delta t^\prime &=& t_2^\prime-t_1^\prime\\
&=&\left[\left(t_2+\frac{|v_\mathrm{pulse}|x_2}{c^2}\right)
-\left(t_1+\frac{|v_\mathrm{pulse}|x_1}{c^2}\right)\right]\Gamma_\mathrm{pulse},
\end{eqnarray} 
resulting in 
\begin{eqnarray}
\label{delay}
\Delta t^\prime=\Gamma_\mathrm{pulse} \left(\Delta
  t-\frac{v_\mathrm{pulse}}{c^2}W/2\right)\quad.
\end{eqnarray}
For protons, the pulse width is much smaller than the gyro radius
and the trajectory is approximately a straight line. The path
length then equals  the pulse width so the time spent inside the pulse
is $\Delta t \approx (W/2)/v_\mathrm{pulse}$. Then Eq.~(\ref{delay})
reduces to $\Delta t^\prime\approx\Delta t /\Gamma_\mathrm{pulse}$. 

The time spent in the pulse
for particles (like intermediate energy electrons) 
with finite curvature trajectories 
is  
\begin{eqnarray}
\label{deltat}
\Delta t =
\Delta t= \alpha_0/\Omega^\mathrm{ext}_\alpha =
\frac{\arcsin{\left(\frac{W}{2R_L^\alpha}\right)}}{\Omega^\mathrm{ext}_\alpha},
\end{eqnarray}
where
$\alpha_0$ is the angle of particle gyration.  

For a proton with $R_L^p>>W$
the path length inside the pulse is 
given by $S_p\approx W$, whereas electrons with
$R_L^e\sim W$ have $S_e>W$. Because both protons and electrons
move with approximately the same speed $\sim c$, 
electrons spend more time inside the pulse. 
The pulse crossing will therefore be observed later for electrons than for
protons. Figure~\ref{fig:5} shows the expected time delays in the
laboratory frame for  the simple model of pulse particle encounter depicted in
Fig.~\ref{fig:2} by using Eqs.~(\ref{delay}) and~(\ref{deltat}).
The expected delays between electrons and protons
agree with measured results from the simulations to within a factor of 2.

\begin{figure}
  \begin{center}
     \resizebox{0.7\hsize}{!}{\epsfig{file=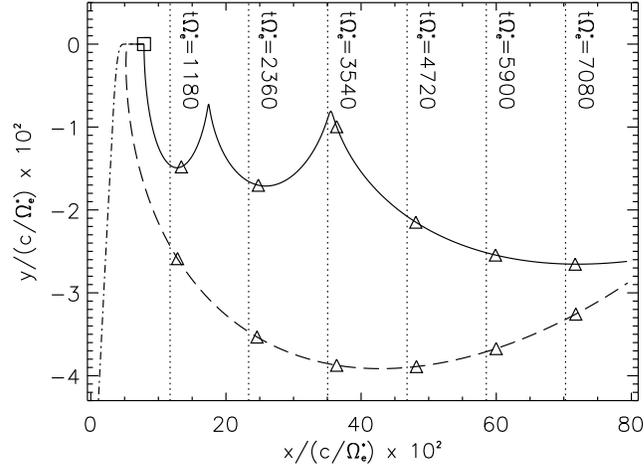}}
  \end{center}
  \caption[]{Electron trajectories for electrons encountering the EM
    pulse at three different initial velocities in the $-x$ direction . 
    The solid line corresponds to
    zero initial velocity, the dashed line to
    $\gamma_e\sim1.6$ and the dash-dotted line to $\gamma_e\sim
    117$. Triangles indicate particle positions at given times along
    the trajectories and dotted vertical lines are the
    positions of the moving pulse. All particles start at the position
    indicated by a square. The vertical dotted line represents the pulse location at the times indicated.} 
  \label{fig:6}
\end{figure}
\subsection{Tracking single electron trajectories}
In the laboratory frame (which is the simulation frame), the
pulse also contains an electric field, 
$\vec{E^\prime}=(0,E_y^\prime,0)$, 
perpendicular to the direction of
expansion and to 
the magnetic field.  The 
electric field strength is given by $ E_y \prime\sim  v_\mathrm{pulse}
B_z^\prime$. Electrons entering the pulse's EM field 
therefore experience an $\vec{E}\times\vec{B}$ - drift in 
the +x-direction. This drift velocity equals  the pulse
velocity so the particles are carried with the 
pulse. 

To track a single particle in PIC
simulations and test for the influence of the drift force, 
we define it as its own species and add it to the 
simulation. By varying the  
initial velocity of the single particle while keeping 
everything else the same, we can  mimic different pulse
velocities and compare the resulting particle trajectories. 
We use an initial
$\intbeta=0.01$ and a temperature of $\intte=\inttp=5\;\mathrm{MeV}$.

The trajectories for electrons with three different initial 
 velocities are displayed in Fig.~\ref{fig:6}. We take the
internal plasma to be a proton-electron plasma.
The fastest electron corresponds to the case of transmission described
in the preceding section. As the electron enters the EM field of the pulse
it is subject to the drift force just discussed and an electric force 
in the y-direction. After a partial 
gyration, the particle leaves the pulse downstream.

The simulation with the electron initially at rest corresponds to 
an cold ambient electron encountering the pulse. The particle
starts gyrating and experiences an $\vec{E}'\times\vec{B}'$ - drift
which drags it along the front of the pulse.

The electron with the intermediate quasi-relativistic velocity 
has a nearly open trajectory, and is not able to perform
a full gyration  during the transit shown.
Although it appears to stay  a constant
distance ahead of the pulse (triangles in
Fig.~\ref{fig:6}), comparing the particle Lorentz factor 
with that
of the pulse reveals that the electrons are actually moving faster than
the pulse but moving toward it.

For the intermediate velocity electron, the peak Lorentz factor
$\gamma^\mathrm{med}_\mathrm{max} \approx 82$ and for the zero initial velocity
electron it is $\gamma^\mathrm{zero}_\mathrm{max} \approx 42$. Both
values peak when the electron velocity is aligned with the pulse velocity. The
pulse Lorentz factor at this time is around
$\gamma^\mathrm{pulse}_\mathrm{max} \approx 12$. 
\section{Long term evolution}
\label{longterm}
Here we take an \ee internal plasma and an \ep external plasma. 
The simulation box size for this simulation
is $L_x=8000\Delta,\;L_y=10\Delta$. For an initial
plasma beta of $\beta^\mathrm{int}_e=0.01$ at a temperature of
$T_e^\mathrm{int}=5\;\mathrm{MeV}$ we can follow the pulse
evolution up to a time of $t\Omega_e^\ast\sim157000$ (compared to 
$t\Omega_e^\ast \sim 10^4$ in the previous section). 
\subsection{Long term energetics}

\begin{figure}
  \begin{center}
     \resizebox{0.7\hsize}{!}{\epsfig{file=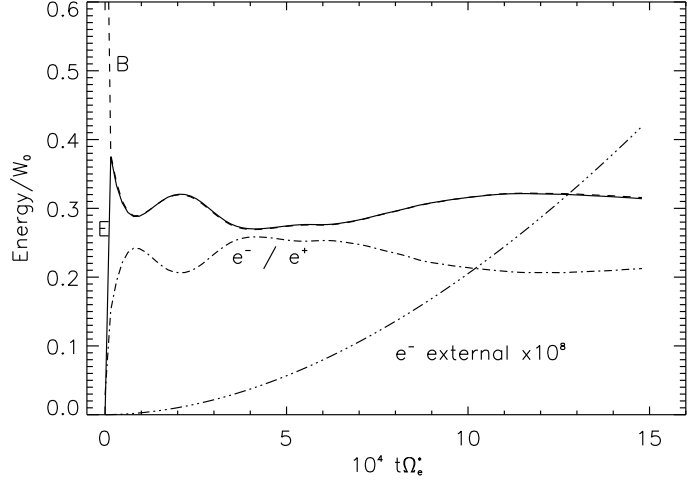}}
  \end{center}
  \caption[]{Energy evolution of the system: total magnetic field
    energy, total electric field energy and total internal electron/positron
    and total external electron energies (enhanced by a factor of
    $10^8$) are shown. All energies are normalized to the initial
    magnetic field energy $M_0$.}
  \label{fig:7}
\end{figure}
The total energy of fields and particle species are
displayed in Fig.~\ref{fig:7}. As described in Ref. ~\cite{liangetal2003},
an initial static magnetic field results in two 
electromagnetic pulses propagating in opposite directions
 into the surrounding volume. Due to the plasma expansion, the
energy in the magnetic field rapidly decreases and 
is transferred to a growing electric field of 
$E_y\sim v_\mathrm{pulse}B_z$
  until an equipartition value
in the field energy is reached. This energy transfer happens in the first
$\sim150\;\gyrotime$. After that, 
the internal particles gain significant energy. 
Internal electrons and positrons experience the same energy increase
and are plotted as one line in Fig.~\ref{fig:7}. 
The external electrons gain energy right from  the start of the
expansion. Due
to the low ambient particle density  (i.e. the ISM), 
its energy content is  small and in order to be visible in
Fig.~\ref{fig:7} we have scaled it by a factor of $10^8$.
The internal particles continuously gain energy until 
$\gyrotime\approx 8000$. At this point the total particle
energy starts to decrease and the electromagnetic field regains
energy. Such behavior has also been observed in Run D of
~\cite{nishimuraliang2004} but the subsequent evolution 
was not followed there.

In the long term evolution seen in
Fig.~\ref{fig:7}, the energy exchange between
internal particles and the electromagnetic field is somewhat
oscillatory in time. 
\begin{figure}
  \begin{center}
     \resizebox{0.7\hsize}{!}{\epsfig{file=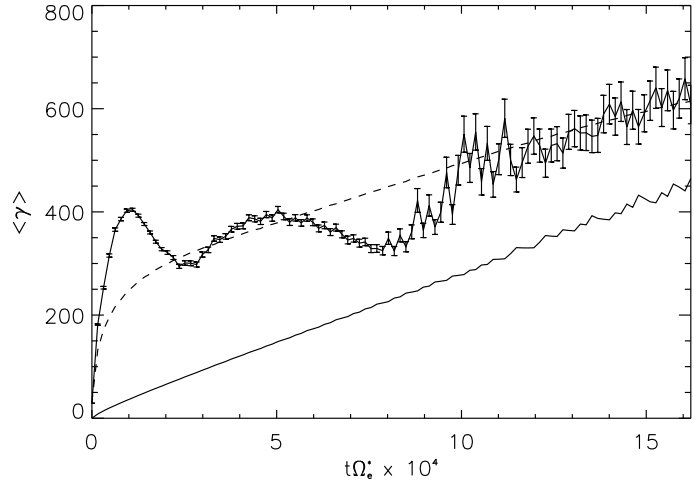}}
  \end{center}
  \caption[]{Mean Lorentz factor $\langle\gamma\rangle$ of the
    internal electrons 
    (upper solid line with error bars) and external electrons (lower
    solid line) vs. time. The dashed line represents the fit of
    Eq.~(\ref{sqrt_law}) to the internal data and the error bars
    are the standard error of the mean. The fit parameters are $f=9.80$
    and $C_0=927.96$.} 
  \label{fig:8}
\end{figure}
Although the total energy of the entire particle population can 
decrease due to particle leakage,
the energy in the trapped particles  still rises~\cite{liangetal2003}.
As a measure of their energization, we show the evolution of the mean Lorentz
factors
\begin{eqnarray}
\label{mean_gamma}
    \langle\gamma(t)\rangle=\frac{\int\limits_\mathrm{pulse}
    N(\gamma(t))\gamma(t)\;d\gamma} 
    {\int\limits_\mathrm{pulse} N(\gamma(t))\;d\gamma}
\end{eqnarray}
of the internal/external electron population as a function of time in
Fig.~\ref{fig:8}. For internal particles, 
the integrations in Eq.~(\ref{mean_gamma}) extend over
all electrons in the leading pulse and for external electrons,
the integration is over a surface layer of thickness $\sim
12\;c/\omega_e$ upstream of the pulse.

\begin{figure}
  \begin{center}
     \resizebox{0.7\hsize}{!}{\epsfig{file=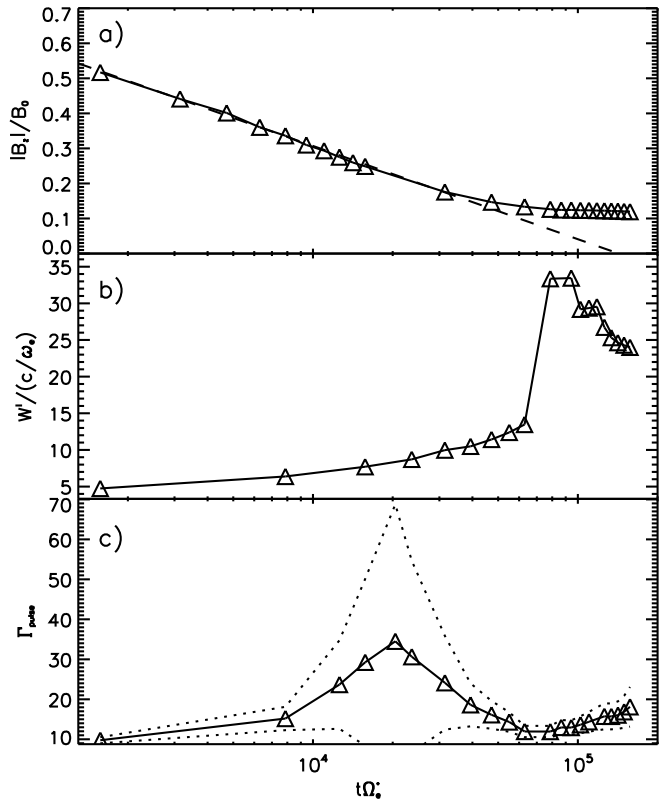}}
  \end{center}
  \caption[]{Pulse evolution in the course of the longterm
    simulation. Panel {\em a)} shows the maximum magnetic field of the leading
    peak (not necessarily the maximum of all peaks in the
    whole pulse). The pulse width $W^\prime$ in the laboratory frame
    is displayed in panel {\em b)} and the pulse Lorentz factor derived from
    the peak position in time is displayed in panel {\em c)}. The
    dotted lines show the error margin.}
  \label{fig:9}
\end{figure}

The mean energy gain for electrons inside the pulse has been derived
analytically to obey ~\cite{liangnishimura2004} 
\begin{eqnarray}
\label{sqrt_law}
   \langle\gamma(t)\rangle=\sqrt{2f\Omega_e(t)t+C_0}\quad,
\end{eqnarray}
where $C_0$ and $f$ are constants that depend on the initial plasma
conditions and $\Omega_e(t)$ is the non-relativistic
instantaneous gyro frequency. Though only co-moving electrons
are assumed in the derivation of Eq.~(\ref{sqrt_law})
$\langle\gamma(t)\rangle$ this is not the pulse Lorentz factor.
The latter is determined by averaging {\it before} squaring, namely
$\Gamma_\mathrm{pulse}=1/\sqrt{1-<\beta_x>^2}$.  This gives comparable
results to the magnetic peak velocity shown in Fig.~\ref{fig:9}c. Our
interpretation of Eq.~(\ref{sqrt_law}) therefore
differs from that of ~\cite{liangnishimura2004}. 

At early times, 
large deviations from the theoretically predicted time dependence are
observed (Figure~\ref{fig:8}). This is due to the assumptions made in the
derivation of Eq.~(\ref{sqrt_law}). 
The fit parameter $f$, which is assumed to be constant in time,
includes the averaged pulse profile. The pulse
shape, however, changes during the evolution of the
system and $f$ is expected to change with time. For example, the deviation 
 from Eq.~(\ref{sqrt_law}) at $t\gyro\approx 8\cdot10^4$ in Fig.~\ref{fig:8} 
is caused by the coalescence of the leading pulse with a trailing pulse,
which suddenly increases the pulse width (see Fig.~\ref{fig:9}b). 

A full analysis of the energy evolution  requires 
consideration of energy losses due to radiation. An estimate of the
synchrotron loss-time yields  
\begin{eqnarray}
  t_\mathrm{syn}\approx 10^{-7}\;\mathrm{s}\left(\frac{100}{\gamma_e} 
  \right)\left(\frac{10^3\;\mathrm{T}}{B_z}\right)^2\quad.
\end{eqnarray} 
Using  $B_z\sim1\;\mathrm{T}$ and assuming a
Lorentz factor $\gamma_e\sim10^4$ yields $t_\mathrm{syn}\approx 10^{-3}\;\mathrm{s}$. Synchrotron
losses are therefore not relevant for the duration of the 
simulation presented herein
and can be neglected. However, we will later consider them in the 
extrapolation to late times.

\subsection{Long term pulse evolution }
The evolution of the pulse is important 
for acceleration  of the external particles. 
The pulse width, peak magnetic field and velocity determine 
whether the pulse will  reflect, trap, or transmit the ambient particles.

The three  key pulse properties are displayed in
Fig.~\ref{fig:9}. Only the leading pulse 
is considered since its values determine the interaction at
the outermost interface with the ambient plasma (the peak magnetic
field plotted is only that of the leading sub-pulse, not the global maximum).
Figure~\ref{fig:9}a shows the decay of the peak magnetic field of the
leading pulse. The dashed line displays the line of a logarithmic decay
of the form $f(t)=1-C_0\log{C_1t}$. The pulse evolution follows 
this formula for 
$\gyrotime\approx3\cdot10^4$. For later times, the peak magnetic field
of the pulse stabilizes and seems to asymptotically approach 
of $|B_z|/B_0\approx 0.12$.

Figure~\ref{fig:9}b shows the time evolution of the
pulse width $W$, defined as the FWHM value. 
The pulse exhibits a continuous broadening until a time 
$\gyrotime\approx 7\cdot10^4$ where a jump occurs. The sudden change
is due to coalescence of a 
trailing sub-pulse with the leading pulse. Predicting the 
longer term pulse width evolution 
is difficult but we can 
estimate that, to order of magnitude, $W\sim 10 \inertial$. This is consistent
with the $\sim\Gamma_\mathrm{pulse}\inertial$ estimate in
~\cite{liangnishimura2004} and the pulse velocities shown in panel
{\em c)} of Fig.~\ref{fig:9}.  

The Lorentz factor of the leading pulse is displayed in
Fig.~\ref{fig:9}c. The pulse velocity has been determined 
by fitting a Gaussian to the uppermost peak where the pulse is closest
to being symmetric. The position of the centroid at the corresponding
times then yields the pulse velocity. Starting at
$\Gamma_\mathrm{pulse}\approx 10$, the pulse reaches a peak 
of $\Gamma_\mathrm{pulse}\approx 35$ and decelerates later to a value of
$\Gamma_\mathrm{pulse}\approx 12$. Toward the end of the simulation,
the pulse gains speed and ends up with 
$\Gamma_\mathrm{pulse}\approx 19$. The low Lorentz factors at times
around $7\cdot 10^4\gyrotime$ derive from the coalescence of the
leading pulse with a trailing pulse (see the jump at the same time in
panel {\em b)} of Fig.~\ref{fig:9}).

So, at the latest times, the peak magnetic field strength is given
by a nearly constant $|B_z|/B_0\approx 0.12$. 
the pulse width seems to remain close to $W^\prime\approx10\inertial$, 
and $\Gamma_\mathrm{pulse}\approx15$.
\section{Radiation spectra}
\label{radiation}
Highly energetic electrons moving in magnetic fields generate
electromagnetic radiation. In the following we derive and discuss the
radiation spectra that are expected to be produced by the DRPA. The
spectra are obtained from a given electron energy
distribution in our simulations and an
emission model, e.g. synchrotron emission. A different approach can be
found in~\cite{koichietal2004} where the emitted radiation is 
self-consistently computed.
\begin{figure}
  \begin{center}
     \resizebox{\hsize}{!}{\epsfig{file=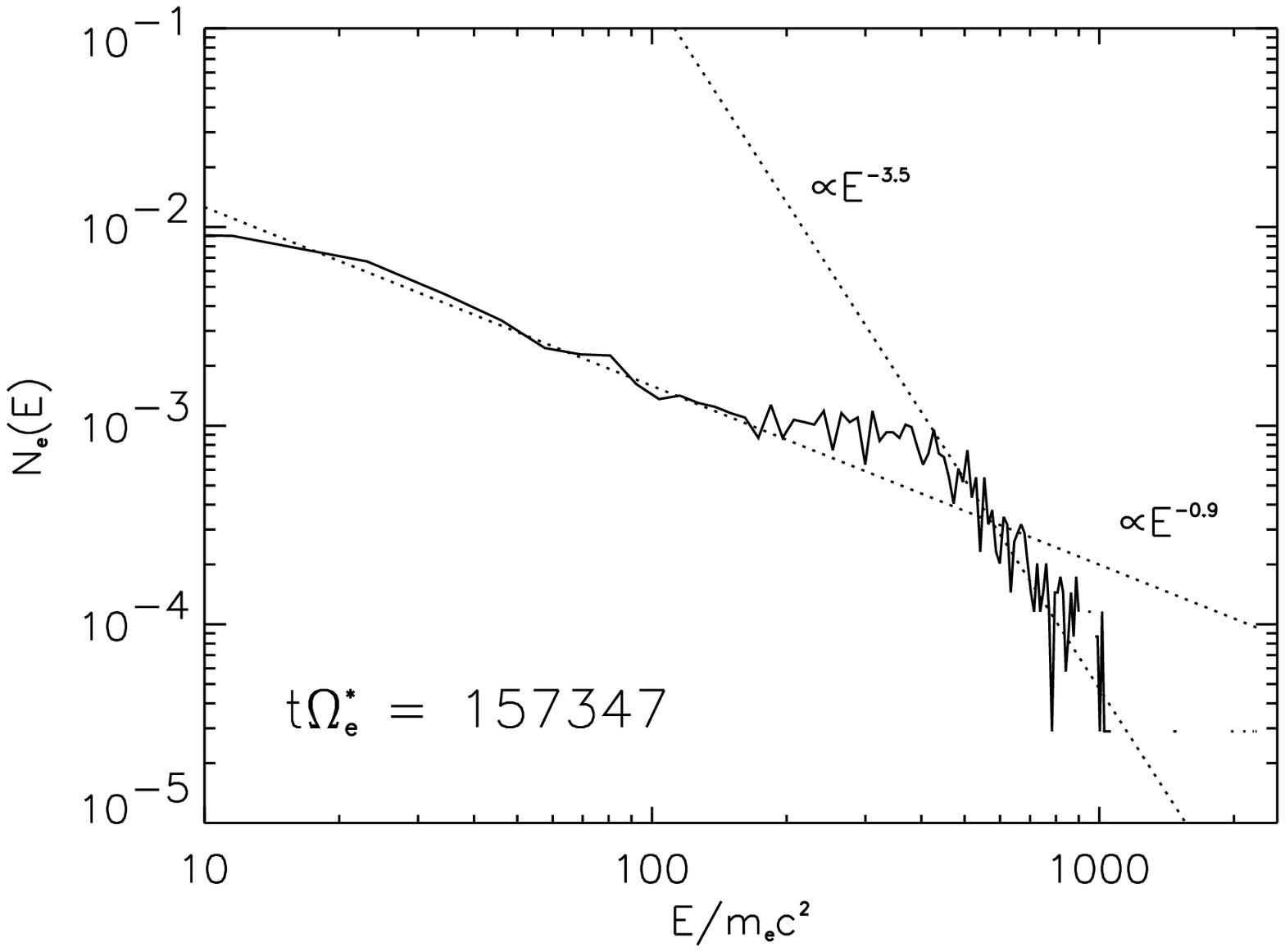}
                           \epsfig{file=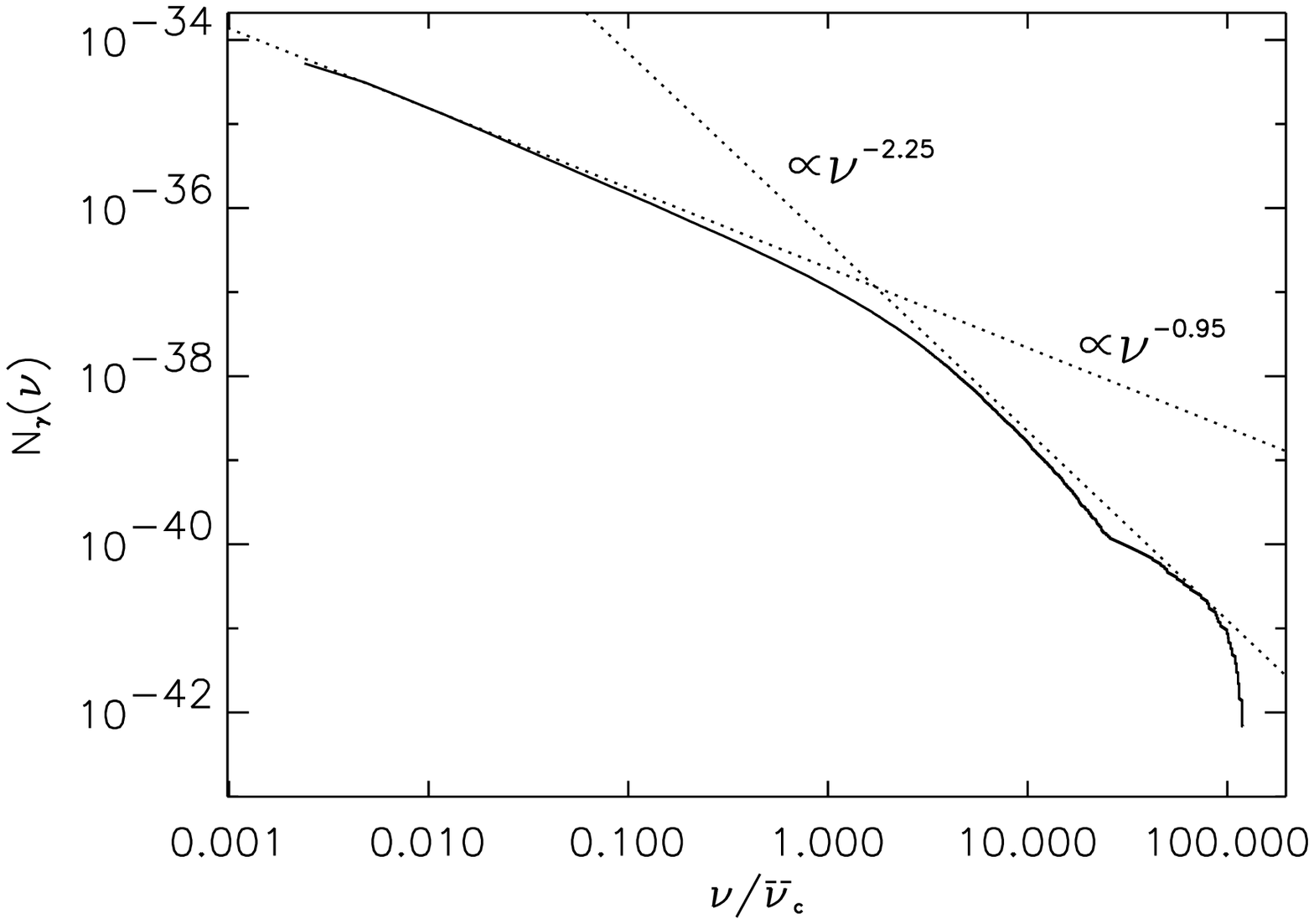}}

  \end{center}
  \caption[]{{\em\bf Left:} The energy spectrum of the internal electrons at a time
    of $\gyrotime = 157347$. Two sections of the spectrum are fit by
    power-laws with slopes $\delta=0.9$ at low energies and
    $\delta=3.5$ at high energies. {\em\bf Right:} Photon number
    density vs. frequency. The frequency is normalized to the 
    critical frequency $\bar{\nu}_c=\bar{\nu}_c(\bar{\gamma},B)$
    (Equation~\ref{nu_sync}) with 
    $\bar{\gamma}=<\gamma_e^\mathrm{int}>$ and $B$ is the peak magnetic
    field in the pulse. The dotted lines display the slopes
    expected from Eq.~(\ref{powerlaw_photon}). The spectrum computed from the
 particle distribution produced by the simulation  is shown by the solid line. }
  \label{fig:10}
\end{figure}
\subsection{Synchrotron vs. jitter}
\label{sync_vs_jitter}
Due to the large electron
Lorentz factors the radiation is beamed into a narrow cone
with opening angle $\Delta\theta\sim 1/\gamma$. Two different types of
radiation can emerge: 1) synchrotron
radiation~\cite{rybickilightman1979} and 2) 'jitter'
radiation~\cite{medvedev2000}. For synchrotron
radiation the characteristic emission frequency is
determined by the particles gyration on a circular orbit. The emitted
radiation beamed in to $\Delta\theta$ is directed toward the
observer only for a short fraction of the gyration. The radiation is
pulsed with a typical duration $\tau_p\sim1/\omega_e$. Thus, the
emission frequency is determined by the particle gyro frequency and
the exact expression is given by (e.g.~\cite{longair1994})
\begin{eqnarray}
\label{nu_sync}
\nu_c^\mathrm{sync}=\frac{3}{2}\nu_e\gamma^2=\frac{3eB\gamma^2}{4\pi m_e}\quad,
\end{eqnarray}
where $\nu_e=\Omega_e/2\pi=eB/2\pi m_e$. 

Jitter radiation occurs if the observer always remains
in the beam of the emission from the radiating particle.
The particle motion can be deflected by
small-scale (smaller than the particle gyro radius) fluctuations in the 
magnetic field, 
resulting in a random scattering of the particle trajectory. But if the
deflections are small, and if the beam angle is larger
than the pitch angle,  the radiation beam always points toward the
observer. Modulation occurs with a frequency determined by the scale
length of the inhomogeneities of the magnetic field and the particle
velocity. The  characteristic emission frequency is  
\begin{eqnarray}
\label{nu_jitter}
\nu_c^\mathrm{jitter}=\frac{\gamma^2 c}{\lambda}\quad,
\end{eqnarray}
where $\lambda$ is the scale length of the magnetic field inhomogeneities.

The characteristic quantity which determines the dominant emission
process is the ratio between the angle of
gyration a particle performs on a scale length of the magnetic field,
$\alpha_0$, and the opening angle of emission $\Delta\theta$,
i.e. $\delta\sim\alpha_0/\Delta\theta$~\cite{medvedev2000}: for
$\delta>1$ the emission is synchrotron and 'jitter'
otherwise. Using  Eq.~(\ref{deltat}), 
the condition for 'jitter' radiation 
is equivalent to 
\begin{eqnarray}
\label{lambda_cond}
\lambda<\frac{c}{\Omega_e}\quad.
\end{eqnarray}
This condition is independent of 
the particle velocity.  
The typical length scale of the magnetic inhomogeneities in the
magnetic pulse, i.e. the sub-pulses due to fractionation, are of the
order of a few $\Gamma_\mathrm{pulse}
c/\omega_e$, hence 
\begin{eqnarray}
\lambda_\mathrm{DRPA}\sim\frac{\Gamma_\mathrm{pulse} c}{\omega_e}=\frac{ c}{\Omega_e}\Gamma_\mathrm{pulse}\left(\frac{\Omega_e}{\omega_e}\right)\quad.
\end{eqnarray}
Since our plasma is magnetically dominated
$\Omega_e/\omega_e>1$ and we find $\lambda_\mathrm{DRPA}>c/\Omega_e$.

We therefore conclude that the dominant radiation process for
electromagnetic radiation from the DRPA is synchrotron emission. This
conclusion  differs from that of ~\cite{liangnishimura2004}
where 'jitter' radiation was assumed.
\subsection{Synchrotron radiation spectra}
\label{synch}
The energy loss $dE/dt$ resulting from synchrotron
radiation is given by (e.g.~\cite{longair1994})
\begin{eqnarray}
\frac{dE}{dt}=\frac{e^2\gamma^4}{6\pi \epsilon_0 c^3}\left(\left(\vec{E}+\vec{v}\times 
    \vec{B}\right)^2-\left(\vec{E}\cdot\vec{v}\right)^2\right)
    \cdot \gamma^2\quad.
\end{eqnarray} 
When $\vec{E}$ is ignored,
the spectral intensity  is 
then 
\begin{eqnarray}
\label{intensity_single}
I_\nu=\frac{\sqrt{3}e^3B\sin{(\alpha_p)}}{4\pi \epsilon_0 m_ec}F(x)\quad
\end{eqnarray}
where
\begin{eqnarray}
F(x)=x\int^\infty_x K_{5/3}(z)\;dz,\quad
\end{eqnarray}
and $K_{5/3}(z)$ the modified Bessel function of order $5/3$. The
argument $x=\nu/\nu_c$ is the ratio of frequency over the critical
frequency (Equation~\ref{nu_sync}). 
The equations can be simplified for 
$\vec{v}\cdot\vec{B}=0$, a reasonable assumption in our case since
$\vec{B}=(0,0,B_z)$, and  particles are mainly energized in 
the x,y-direction.

The energy loss for a single electron averaged over all pitch-angles
is then given 
by~\cite{longair1994}
\begin{eqnarray}
\label{loss_rate}
-\left(\frac{dE}{dt}\right)=\frac{4}{3}\sigma_T c U_B\beta^2\gamma^2,
\end{eqnarray}
where $\sigma_T$ is the Thompson cross section, and $U_B=B^2/2\mu_0$ is
the magnetic energy density. In the relativistic limit
$\gamma\gg1$ the normalized velocity $\beta=v/c$
can be set to unity. 

The emission spectra are computed by summing
Eq.~(\ref{intensity_single}) over all particles. The resulting spectra
are compared to the analytic model spectra for particle
energy distribution functions of the form
\begin{eqnarray}
\label{powerlaw_ptcl}
N_e(E)\;dE=C_e\cdot E^{-\delta}\;dE,\quad
\end{eqnarray}  
where $\delta$ is typically positive, and takes constant values
over a finite energy range. 
This  power law  translates into photon number density 
of the form ~\cite{rybickilightman1979}
\begin{eqnarray}
\label{powerlaw_photon}
  N_\gamma(\nu)=C_\gamma\cdot\nu^{-\tau},\quad
\end{eqnarray}
where $\tau$ is given by $\tau=(s+1)/2$.
\subsection{Radiation from internal particles}
At the latest times in our simulations, the internal particle energy spectrum 
can be approximated by a double power-law.  
Here we will assume isotropic particle
distributions.
This is satisfied for the internal electrons: Although the 
trapped electrons get anisotropically accelerated, the pulse
fractionates at late times,
randomizing the magnetic field and thus  the particle
velocities. At the end of the simulation run, when the photon
spectrum is computed, the particle distribution has been roughly isotropized. 
The internal electron 
energy spectrum at the end of the long term simulation run is
shown in Fig.~\ref{fig:10}. The region below
$E\approx200\;{m_ec^2}$ is well fit by a power-law with
spectral index $\delta\approx0.9$ while for 
$E\geqslant400\;{m_ec^2}$ we find 
$\delta\approx3.5$. 
An excess of particles is observed at the transition 
which might be interpreted as another power-law with very small slope.

According to Eq.~(\ref{powerlaw_photon}), a double power-law dependence
is also expected in the photon number spectrum with index 
$\tau\approx0.95$ 
at lower energies and $\tau\approx 2.25$ above the break energy. 
Summing the contributions from all internal electrons 
using  Eqs.~(\ref{intensity_single}) and~(\ref{nu_sync}) yields
the photon spectrum shown in Fig.~\ref{fig:10}. The analytically 
predicted slopes are shown as dotted lines. They agree 
well with the actual spectra computed from the particle distributions
produced by  the simulation, although 
at lower frequencies, the spectrum exhibits a slope of $\tau\approx
1.05$ rather than the predicted $\tau\approx0.95$. 
At higher energies, $\tau\approx 2.25$ as predicted. 
At high energies there is a hump in the spectrum which comes 
from a fast population of electrons visible as small dots around an
energy of $E\sim 2000\;\mathrm{m_ec^2}$ in Fig.~\ref{fig:10}. These
particles  do not fit into the
high energy part of the power-law. Only very few particles
reach the highest energies and due to  limited statistics it is
difficult to make conclusive statements about them.  
\begin{figure}
  \begin{center}
     \resizebox{\hsize}{!}{\epsfig{file=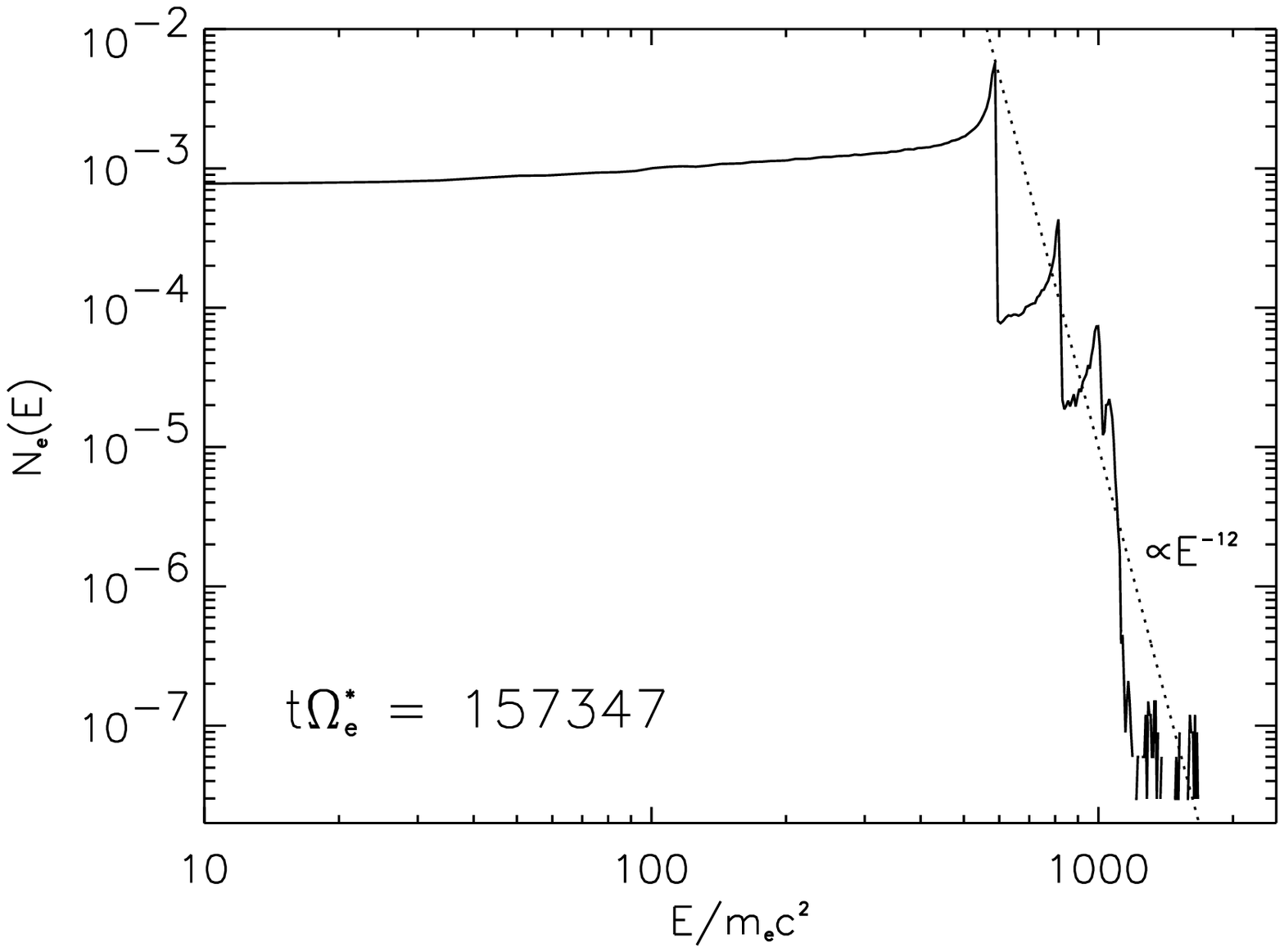}
                              \epsfig{file=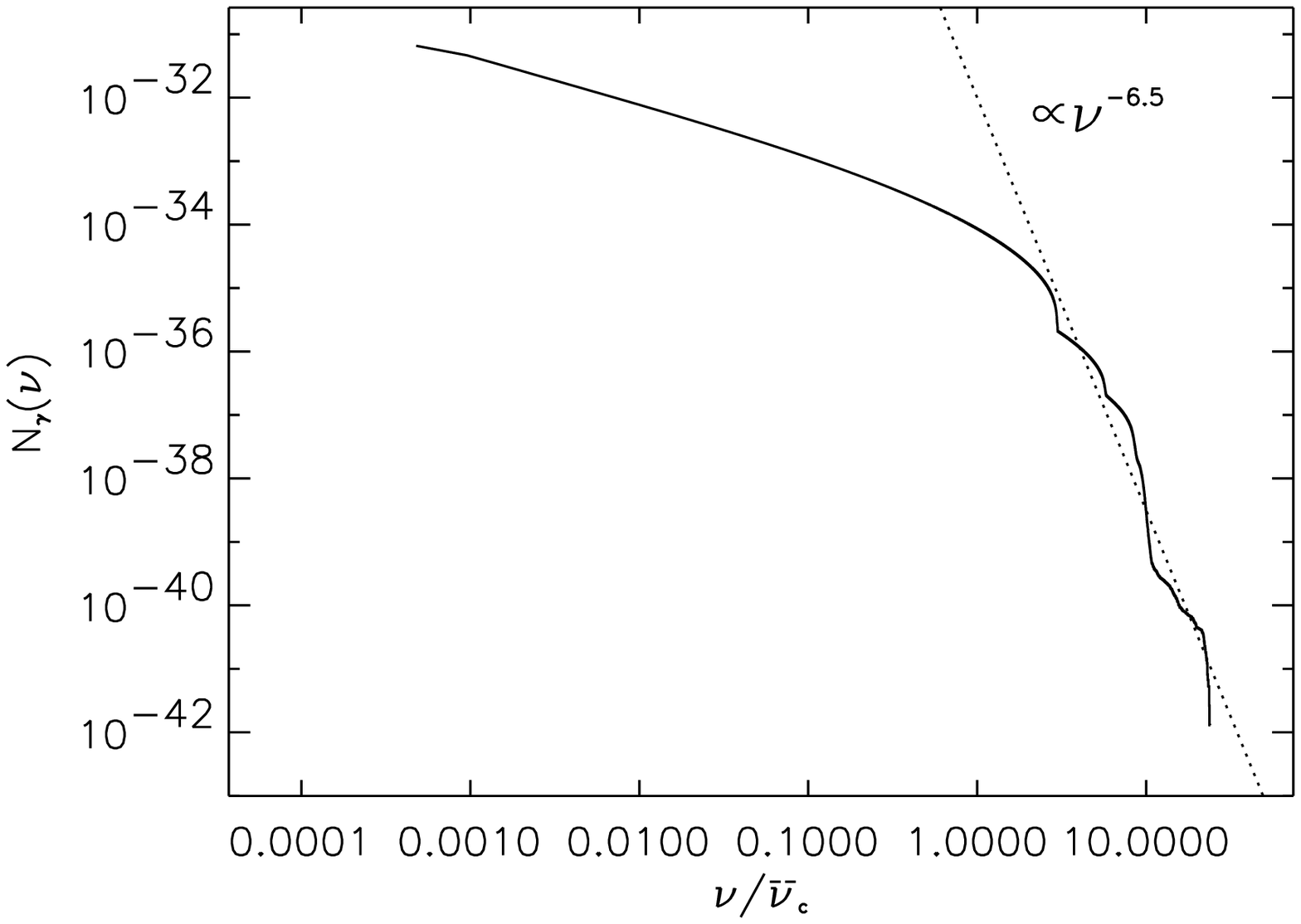}}
  \end{center}
  \caption[]{{\em\bf Left:} The energy distribution of the external electrons at a
    time of $\gyrotime = 157347$. The dotted is line 
    is the slope in the particle distribution that
    corresponds to the slope that fits  the photon spectrum in the
    right panel. {\em\bf Right:} Photon spectrum produced by the external electrons in a
    thin layer ahead of the pulse. The x-axis normalization is the
    same as described in the caption of Fig.~\ref{fig:10}.} 
  \label{fig:12}
\end{figure}
\subsection{Saturation and termination of internal electron acceleration}
\label{saturation}
Energy is continuously transfered from the pulse magnetic field 
into the internal electrons. Some of the energy
is radiated away  by the electrons. Here we estimate the associated time
scale and  energies for which these processes balance.

A net energy build-up in the internal electrons only occurs
if the energy gain rate exceeds the loss. But 
the system will eventually reach a steady-state where the energy input
balances the energy output by radiation. Knowing the mean acceleration
rates of the DRPA and assuming synchrotron emission to be the dominant
loss, the time and energy scale to reach saturation
can be estimated. 
The mean acceleration rate of the internal electrons can be obtained via
$\frac{d\langle E\rangle}{dt}_{ac}=m_ec^2\frac{d\langle\gamma\rangle}{dt}$.
Using Eq.~(\ref{sqrt_law}) and values of $f$ and $C_0$
obtained from fits to the simulation data in this expression
and equating to the synchrotron loss-rate from
Eq.~(\ref{loss_rate}). Comparison of the mean $\gamma^2$ and the square
of the mean $\gamma$ in the simulated electron distribution yields  
$\langle\gamma^2\rangle\approx\langle\gamma\rangle^2$ within a factor
of $1.5$. We can therefore write
\begin{eqnarray}
\label{balance_time}
\gyro t_\mathrm{bal} = \frac{1}{2f\rho}\left[\left(\frac{3}{2}\frac{\mu_0 e c
      f}{\sigma_T\rho B_0}\right)^\frac{2}{3}-C_0\right],
\end{eqnarray}
where $\rho\equiv|B_z(t)|/B_0$, assumed to be a constant $(\sim 0.12)$
for times $t\gyro\geqslant 10^5$ (see
Fig.~\ref{fig:9}a). Equation~(\ref{balance_time}) gives the time-scale
for saturation of the DRPA. 
Using our initial simulation parameters, we find a saturation time of
$t_\mathrm{bal}\approx 0.0005\;\mathrm{s}$. The mean electron energy
reached at $t_\mathrm{bal}$ is $\langle\gamma(t_\mathrm{bal})\rangle\approx
3\cdot10^4$ resulting in a synchrotron photon energy 
$E_\gamma\approx 100\;\mathrm{keV}$. This energy can be corrected to
larger values due to underestimating $\langle\gamma^2\rangle$ in
Eq.~(\ref{balance_time}). 

\subsection{Radiation from external particles}
The energy distribution function of the external electrons piled up
ahead of the pulse is primarily shaped by the particles' 
gyration as they encounter the pulse magnetic field. 
As seen in Fig.~\ref{fig:12}, the 
collective gyration  creates strong spikes in the
energy distribution. 

Summing the contributions to synchrotron emission from all 
external particles using Eqs.~(\ref{intensity_single})
and~(\ref{nu_sync}) yields 
the photon spectrum displayed in Fig.~\ref{fig:12}. The resulting
slope at high energies is $\tau = 6.5$.
The corresponding slope in the particle energy distribution 
according to Eqs.~(\ref{powerlaw_ptcl})
and~(\ref{powerlaw_photon}) is shown as dotted line in Fig.~\ref{fig:12}.
Such a steep slope is not observed in any phase of the GRB
emission, but our simulations cover at most only a microsecond
of a real burst. At such times, the density of the external particles
accumulated is too low to produce any observable radiation. Only
after the prompt emission would the accelerated external population
contribute to observable emission. The external particle spectrum is
likely to  change significantly  by then. 

Although the energy distribution of the electrons exhibits strong
population inversions, no instabilities are observed during the
simulation time-span. A two-stream like instability might be
expected due to the gyrating beam creating many counter streaming
micro-beams in close vicinity to the pulse upstream. 
The reason the absence of instabilities 
lies in the highly relativistic nature 
of the system.Ref.  \cite{medvedevloeb1999}  calculates the maximum
growth rate for e.g. the magnetic Weibel instability~\cite{weibel1959}. 
For a small thermal spread compared to the
bulk motion of the beams ($\Gamma_b\approx\Gamma_{b\parallel}$ and
$\Gamma_{b\perp} \ll\Gamma_b$, where $\Gamma_b$ is the total Lorentz
factor of the streaming electrons and $\Gamma_{b\parallel,\perp}$ are
the parallel and perpendicular components), the maximum growth
$g_\mathrm{max}$ 
rate is given by 
\begin{eqnarray}
  g_\mathrm{max}^2\approx\frac{\omega_e^2}{\Gamma_b}\left(1-2\sqrt{2}
  \frac{\Gamma_{b\perp}}{\Gamma_b}\right)\quad.
\end{eqnarray}
For large $\Gamma_b$ the growth rate becomes small.
\subsection{Further discussion of the photon spectra and limitations of our calcuations}
Using the particle energy distribution functions from the simulations
to compute the synchrotron spectra requires several
approximations. As 
pointed out in Sect.~\ref{radiation} it is assumed that
$\vec{v}_e\perp\vec{B}$, which allows us to ignore the angular
correction factor $\sin{(\alpha_p)}$ in
Eq.~(\ref{intensity_single}). Also, the magnetic field in the emission region
is assumed to be constant and its magnitude is given by the peak value
in the pulse.  

To compare the analytical results in
Eq.~(\ref{powerlaw_photon}) with the simulations, 
we also assumed the  particle velocity distribution to be  isotropic
which can be justified for internal electrons, as discussed earlier.
We therefore observe a good agreement between the
analytically obtained results for the internal power-law index in
Eq.~(\ref{powerlaw_photon}) and the power-law index obtained from
numerically summing the contribution of each electron (see
Fig.~\ref{fig:10}).    
External reflected particles, however, are not isotropized on the time
scales covered by the simulations. The external electrons do not reach the
region of randomized magnetic fields and therefore maintain their 
anisotropy beyond the end of our simulation. The dotted lines in
Figs.~\ref{fig:12} and~\ref{fig:12} are therefore plotted only to
guide the eye and do not have the same robustness as for internal
particles. 

Do the photon spectra remain stable at late times? 
For internal electrons, the simulations suggests 
little change when the radiation
loss is unimportant.  This follows because
 the internal electron power law slopes displayed in Fig.~\ref{fig:10} do not
change during the last third of our simulation run.
The spectrum from the external electrons exhibits a
power-law index at high energies of $\tau\approx-6.5$ which is rather
steep.  But even at the end of the simulation, 
the density of the external particles is too low to produce
observable radiation.  
In contrast to the photon spectrum from internal electrons, 
that from the external electrons will likely change with time:
The external particles in the rest frame of
the pulse penetrate the region of magnetic field as a relativistic
cold beam. The collective
gyration of these particles produces spikes in the electron energy
distribution seen in Fig.~\ref{fig:12}. 
But the number of spikes increases 
as more material enters the pulse and eventually 
fill in and smooth out the distribution.
In addition, the beam will not remain cold after penetrating the
magnetic pulse but will  be thermalized under the influence of
the increasingly fractionated and randomized magnetic field. 
Finally, the electron spikes 
represent population inversions in the energy spectrum and one might
expect them to act as sources of free energy for plasma
instabilities, e.g. maser-like instabilities.
But highly relativistic systems tend to
stabilize as most particles have $v\sim c$. Only after the pulse slows
via  interaction with surrounding material, 
would such instabilities become important. These time-scales
are beyond those of our simulations and so 
the asymptotic external particle  synchrotron photon spectrum cannot be 
predicted from our simulation results.

The presence of this external radiation component  is a
direct result of the pulse interaction with the 
ambient medium. 
Due to the low density, the time needed
to accumulate enough material to produce
observable radiation implies that a delayed emission component arises with
respect to the prompt emission from the internal particles. 

We conclude this section by summarizing the 
important limitations of our radiation calculations.
We have used the particle distribution function from the simulations
that we then use a posteriori to calculate the emission spectra.
The particle distribution function can be directly inferred from the 
PIC calculations because, as stated earlier, the charge to mass ratio
of superparticles and particles are the same. 
Although our calculated spectral break should survive, 
the actual normalization of the 
radiation flux cannot be accurately determined because the number of photons
does depend on the actual particle number.
Another key issue  is that radiation reaction terms are not
included in our present work.
A complete treatment of particle acceleration and radiation 
should dynamically couple the radiation reaction  to the particles
and fields each time step. This is beyond the scope of the present work
and has been addressed by Noguchi et al. (2005). The effect seems to be to 
lower the total  energy in the radiation.  However, all existing
simulations presently do not last long enough to know exactly how 
these results extend to the much longer  
time scales needed for direct  application to astrophysical
sources.  That remains for future work.

\section{The pulses as electron filters}
\label{filter}
As discussed above, if the pulse width
exceeds the proton gyro radius, both, protons and electrons are
reflected. This is the case of 
~\cite{smolskyusov1996}. If the pulse width is smaller than the proton
gyro radius, a  critical value of the initial internal plasma
$\beta_e^\mathrm{trans}$ divides
the parameter space into two regions: one 
in which external electrons are trapped by the pulse 
and one in which the external electrons pass through. The critical value of
$\beta_e^\mathrm{trans}$ depends on the  
temperature of the internal plasma but not its density.
The simulation parameters are such that the
external protons always pass through the pulse unimpeded.
That electrons can be trapped but
protons are always transmitted, is important
in determining the deceleration rate of the pulses.
The pulses can act as  filters that accumulate the ambient 
electrons while allowing the ambient protons to pass through. 
Such filtering pulses 
would then decelerate more slowly than standard ``snow plow'' hydrodynamic
outflows in astrophysics that accumulate both electrons and protons.

The average magnetic field in
the 
pulse can be 
described as a luminosity, i.e. $(B^2/2\mu_0)4\pi R^2 c = L$. 
Balancing the mechanical or electromagnetic outflow luminosity
with the inertia of the ambient matter gives
\begin{eqnarray}
\frac{L}{\Gamma_\mathrm{pulse}^2}=4\pi
R^2\;c\;\Gamma_\mathrm{pulse} n\;\Gamma_\mathrm{pulse} m c^2, 
\end{eqnarray}
where $n$ is the density of the piled-up ambient
medium in the lab frame, $m$ is the constituent particle mass,
$\Gamma_\mathrm{pulse}$ is the 
flow Lorentz factor of the pulse and $R$ is the distance from the
source. This yields  
\begin{eqnarray}
R=\frac{1}{\Gamma_\mathrm{pulse}^2}\sqrt{\frac{L_\mathrm{obs}}{
    4\pi\;m\;n\;c^3}}\quad.  
\end{eqnarray}
If $m$ of the piled-up material
is reduced by $m_p/m_e\sim 1836$, the distance the pulse of a given
$\Gamma_\mathrm{pulse}$ propagates into the ambient medium is increased by a
factor of $\sim 40$. 

\subsection{Electron filter survives  discharge}
How long the electron filtering is maintained
depends on how long it takes for the electric field induced by the
resulting charge separation  to be strong
enough to capture the protons.

In the pulse frame,
the equation of motion for a single proton in the electric
field behind the pulse is given by
\begin{eqnarray}
m_p \frac{d v_p \Gamma}{dt} = -e E
=-(e^2/2\epsilon_0) (n_e(t-r/v_p) w - n_0 r(t)),
\label{26}
\end{eqnarray}
for ${w}\le r\le ct,$
where $w$ is the width of the electron layer, $r$ is
the distance of the test-proton behind the pulse, $n_e$ is the electron
density accumulated in the pulse and $n_0$ is the ambient
proton number density. The second term on the right of the
expression for $E$ is
due to charge screening by the ambient protons having crossed the
pulse unperturbed and filling the downstream region uniformly at a
number density $n_0$.  The retarded time appears in the contribution
to $E$ from the electrons in the pulse because its accumulation
of electrons takes a time $r/v_p$ to propagate to the proton, where $r$ is
measured from the pulse.
Using the fact that the total excess electron density at the pulse
must equal the proton density left behind
$n_e(t)w = n_0 v_p t$.

If we define $t_0$ as the time at which the given
proton crosses the pulse, so  $t-t_0= r/v_p$.
We now assume that $v_p\sim c$ for the regime of our calculation.
Then the appearance of the retarded time above
implies that the electro-static field at the position of the
proton at time $t$ is determined by the electron charge density
$n_e(t_0)$. Equation~(\ref{26}) becomes 
\begin{eqnarray}
\frac{d\Gamma}{dt} \simeq - \frac{e^2}{2 \epsilon_0 m_p c}n_0(ct_0-r(t)).
\label{27}
\end{eqnarray}
While $\Gamma$ can change significantly, $v_p$
remains close to $c$. We therefore invoke $r(t)\simeq v_p(t-t_0)$
and assume
a constant $v_p\simeq c$ for the present calculation.
Inserting this expression into (\ref{27}) yields
\begin{eqnarray}
\frac{d\Gamma}{dt} = - \frac{e^2}{2 \epsilon_0 m_p }n_0(2t_0-t)).
\end{eqnarray}
Integrating this from $t_0$ to $t$ gives
\begin{eqnarray}
\Gamma(t)= \Gamma(t_0)-
\frac{e^2 n_0}{2\epsilon_0 m_p
  }\left[2t_0(t-t_0)-\frac{1}{2}(t^2-t_0^2)\right].
\end{eqnarray}
We can then solve for the time $t=\tau$ when
$\Gamma$ falls to $\Gamma(t_0)/2$. The
physical solution ($\tau > t_0$) is given by
\begin{eqnarray}
\label{decc_time}
\tau(t_0)\simeq 2t_0 +
t_0\left(1+\frac{\Gamma_0}{t_0^2}\frac{2\epsilon_0 m_p}{e^2 n_0} 
\right)^{1/2}\geq 3t_0.
\end{eqnarray}
To use this
result in comparing time scales for
the evolution of $\Gamma$ with models without proton leakage,
we must transform this time scale
back to the lab frame. Since we consider the time for $\Gamma_\mathrm{pulse}$ to fall
only by a factor of 2, in the lab frame,
$\tau^\prime \ge \Gamma_0 \tau/2$ and
$t_0^\prime \le \Gamma_0 t_0$, so that
$\tau^\prime\ge 1.5 t_0^\prime$

That  $\tau'>1.5t_0'$ is important.
This implies that proton drag
does not  signficantly reduce the time scale for
the relativistic electron dominated pulse to slow down
compared to the case in which the protons are ignored. Even at the low
limit that is given by $\tau^\prime=1.5t_0$, the drag exerted on the
pulse is only produced by a fraction of the protons.
In this case the pulse can be considered to propagate with no
additional drag during the last 33\% 
of the propagation: only protons that
crossed the pulse at times up to $t_0^\prime=\tau^\prime/1.5$ 
can influence the pulse.
\section{Application to GRB}
\label{application}
In the following section we address several aspects of the DRPA in the
context of a PFDO model for GRBs.
 
In order to make observable predictions, we 
have to extrapolate our simulation results to times beyond the reach of the
simulations. Such extensions bear a number of
pitfalls but presently offer 
the only way to connect the early time micro physics
with the observations in the absence of long duration simulations.

Equation~(\ref{sqrt_law}) allows us to use the fit parameters obtained
in Sec.~\ref{longterm} (dashed line in 
Fig.~\ref{fig:8}) to estimate the time-scale at which the 
characteristic frequency of the 
emitted radiation from the internal electrons reaches a typical
value for GRB. Using Eq.~(\ref{sqrt_law}), 
and assuming that asymptotically $|B_z|/B_0\approx0.12$
(Fig.~\ref{fig:9}a), the estimated time needed to reach a typical
photon energy of $E_r$ can be calculated from 
\begin{eqnarray}
\Omega^\ast_e t_r=\frac{|B_0|}{B_z}\frac{\gamma_r^2-C_0}{2f}\quad,
\end{eqnarray}
where  $\gamma_r$ is determined by
Eq.~(\ref{nu_sync}), $E_r$ and  $B_0$. $B_0$ here is not the surface
magnetic field of the GRB progenitor but rather the magnetic field in
the region where the Poynting flux emerges, which can be farther away
from the GRB source and can therefore be smaller. For
$E_r= 500\;\mathrm{keV}=h\nu_{r}$  and 
$B_0=4.473\;\mathrm{T}$, $t_r\approx 0.003 \;\mathrm{s}$ which is
consistent with reaching this energy before observed prompt GRB time scales
demand them.


GRB prompt emission seems to exhibit a double
power law emission spectra  ~\cite{bandetal1993}
as observed by BATSE~\cite{bandetal1993,preeceetal2000}.
Remarkably, the photon power law indices obtained from our simulations, 
$\tau\sim 1.05$ at low and $\tau\sim 2.25$ at high energies, are in excellent
agreement with those  observed ~\cite{preeceetal2000}. 

That being said, 
the actual power-law break energy cannot be determined from our simulations
since the actual photon spectra of observed GRBs 
are at much higher energies than reached during the duration of our 
simulations.  Our maximum simulation time ($\sim 10^{-6}
(B_0/1\mathrm{T})^{-1}\mathrm{sec}$)  
is too small for the electrons to reach the energies at which they would
radiate in the gamma-ray range. In order to reach these energies, 
simulations times $>10^{-6}$s are needed. 
We are presently restricted to discussions of the shape rather than
true fluxes and energies. From the considerations on saturation of the
DRPA we know that the energies of the emitted radiation at saturation 
are of order $\sim100\;\mathrm{keV}$, consistent to within an order of 
magnitude with that of the prompt GRB emission.

In addition to the internal accelerated electrons which are
responsible for the prompt gamma-ray emission, the 
external electrons trapped from the ambient medium
would radiate in a later phase. To estimate the onset of their
observable contribution, we must estimate 
the time evolution of their 
density increase in the surface layer upstream of the pulse.
We assume that this layer is thin compared to the pulse
travel distance  (thin-shell approximation). 
In a layer of thickness $\delta r \equiv \delta 
R/R_0 \ll r(t)\equiv R(t)/R_0$, where $R_0$ is the initial 
pulse size, the density increase after a time $t$ is then 
\begin{eqnarray}
\label{density_increase}
\frac{n_\mathrm{shell}(t)}{n_\mathrm{ext}^0}=
  \frac{1}{3}\frac{r(t)^{3}-1}{r(t)^{2}\delta r}\quad.  
\end{eqnarray}
This is valid for  spherical expansion, but since 
the actual situation corresponds to  a 
spherical expansion in a beam of very small opening angle, 
the slab geometry of our simulations offers an acceptable
correspondence. 
Taking $\delta R=100c/\omega_e$ and extending this to a time of 
 $t=10\;\mathrm{s}$ yields a density increase by a factor 
$n_\mathrm{shell}/n_\mathrm{ext}^0\approx1.7\cdot10^9$. Assuming an
initial external plasma density $n_\mathrm{ext}^0=1\cdot10^6\mathrm{m^{-3}}$, 
corresponding to a typical ISM,  $n_\mathrm{shell}\approx
1.7\cdot10^{15}\mathrm{m^{-3}}$ at this time. 


The choice of $\delta r $ corresponds to observations in our
simulations. Increasing its value will decrease the density growth,
resulting in a further delay of the emission with respect to the
prompt GRB emission. Also depending on $n_{ext}^0$ and the pulse 
velocity, this radiation could threfore produce a second peak in the prompt
emission or coincide with the GRB afterglow ~\cite{piran1999}.

\section{Summary}
\label{summary}
The present work focuses on the
propagation of a Poynting flux dominated relativistic 
plasma outflow (slab) with an ambient medium.
We find that the magnetic outflow breaks up into
sub-pulses, consistent with that of the DRPA mechanism~\cite{liangetal2003}.
We study the interaction of the leading pulse with the ambient medium.
The interaction can now be categorized into three regimes: (a) all
external particles are reflected (nearly equivalent to ``trapped''
for relativistic propagation) by the pulse, (b) all
external particles are transmitted, and (c) electrons are reflected but
not ions. 

In all three regimes, the internal particles 
originating from within the initial expanding
plasma (electrons and positrons) 
are always trapped and accelerated. 
External leptons are only trapped in regimes (a) and (c). 
Regime (a), which is realized when the pulse width is much 
larger than the proton gyro-radius, is equivalent to the case discussed in the
literature by~\cite{smolskyusov1996}. In contrast, the focus of the
present work is on  regime (c), which 
gives rise to a new population of energetic electrons. 

For regime (c) the pulse acts as a semi-permeable membrane
that filters out electrons from the ambient medium whilst allowing 
the transmission of ions. 

The different behavior of ions and electrons leads to an important
difference in the drag force excerted on the pulse by the piled up
matter: instead of being immediately felt, the inertia of the protons
acts on the pulse with a delay via electrostatic interaction between
piled up electrons and transmitted protons. In terms of the
propagation time $t_0^\prime$ of the pulse, only protons that have crossed the
pulse at times up to $t_0^\prime=\tau^\prime/1.5$ can influence the pulse.

In addition to these findings, we  have analyzed the
particle energy distributions, acceleration rates and synchrotron
spectra of the external and internal electrons. We were able to extend the   
simulation times to $t\approx 160000\;\gyrotime$, allowing the study of  
significantly longer term evolution than in previous
work. We find:

\noindent $\bullet$
External electrons pile up in a thin shell at the
leading flank of the pulse and are accelerated at a
rate similar to that of the internal electrons. A fractionally 
signficant amount of radiation (compared to that from internal electrons) 
from these piled-up external electrons would occur late 
stage of the pulse evolution, only when their 
density has grown large enough. 

\noindent $\bullet$
The total electron/positron energy quasi-oscillates in exchange with the
electromagnetic field energy. However, the Lorentz factor of the
pulse continues to grow at the analytically 
predicted rate throughout the timespan covered by our simulations, as
electrons are leaked. 

\noindent $\bullet$ The internal particles 
develop a double power-law energy spectrum 
with a clearly defined break energy. The resulting synchrotron
photon spectrum is in remarkable
agreement with observations of GRBs~\cite{bandetal1993,preeceetal2000},
though the small time-scales of our simulations prevent us from presently 
uncovering any 
additional physics that might modify this spectrum at the later times
needed to realistically compare with actual GRB observations, 
This accelerator will saturate at times of order
$0.5\;\mathrm{ms}$ due to increasing energy loss via synchrotron
emission. The characteristic particle Lorentz factor of a particle whose
radiation loss balances its acceleration 
lies at $\sim 100\;\mathrm{keV}$. In the context of GRBs this energy
lies within an order of magnitude of the characteristic photon energy
of $\sim 500\;\mathrm{keV}$.

\noindent $\bullet$
While the initial configuration is strongly magnetically dominated,
at late times in the simulations, at least 1/2 of the energy is
convered into accelerated electrons.  While the simulations here
represent the physics occurring only at the very front of what would be a
much larger scale extended magnetic tower propagating into an ambient
medium, the results show  that the conversion at the front of the
tower is extremely efficient, even without relativistic shocks.

It is now essential to extend the simulations to later times that
allow the analysis and comparison of prompt and delayed emission in
the course of the DRPA. The goal is to directly connect the initial
expansion with the emission of prompt radiation at later times.
Our present extrapolations can then be tested. One way to reduce 
the computational effort is to use a 
moving window that allows to follow the pulse and simulate only
its closer environment. This approach is currently being implemented.

\section*{Acknowledgments}
We thank A. Frank for use of the ORDA linux cluster
and acknowledge support from NSF grant AST-0406799.
GP acknowledges support from the Swiss National Science Foundation.
EGB acknowledges additional support from 
the Isaac Newton Institute for Mathematical
Sciences, Cambridge UK, and  the KITP of UCSB, where this
research was supported in part by NSF Grant PHY-9907949.

\section*{References}


\label{lastpage}
\end{document}